\documentclass[aps,pra,twocolumn,superscriptaddress,nofootinbib,a4paper,floatfix]{revtex4-2}

\usepackage[pdftex]{graphicx}
\usepackage{amsmath}
\usepackage{verbatim}
\usepackage{xcolor}
\usepackage{bm}
\usepackage{txfonts}
\usepackage{siunitx}
\usepackage[colorlinks=true,allcolors=blue]{hyperref}
\usepackage{mathtools}

\begin{document}

\title{Mechanical overtone frequency combs}

\author{Matthijs H.\ J.\ de Jong}
\affiliation{Department of Precision and Microsystems Engineering, Delft University of Technology, Mekelweg 2, 2628CD Delft, The Netherlands}
\affiliation{Kavli Institute of Nanoscience, Department of Quantum Nanoscience, Delft University of Technology, Lorentzweg 1, 2628CJ Delft, The Netherlands}

\author{Adarsh Ganesan}
\affiliation{Ahmedabad University, Ahmedabad, Gujarat 380009, India}
\affiliation{National Institute of Standards and Technology, Gaithersburg, Maryland 20899, USA}

\author{Andrea Cupertino}
\affiliation{Department of Precision and Microsystems Engineering, Delft University of Technology, Mekelweg 2, 2628CD Delft, The Netherlands}

\author{Simon Gröblacher}
\affiliation{Kavli Institute of Nanoscience, Department of Quantum Nanoscience, Delft University of Technology, Lorentzweg 1, 2628CJ Delft, The Netherlands}

\author{Richard A.\ Norte}
\affiliation{Department of Precision and Microsystems Engineering, Delft University of Technology, Mekelweg 2, 2628CD Delft, The Netherlands}
\affiliation{Kavli Institute of Nanoscience, Department of Quantum Nanoscience, Delft University of Technology, Lorentzweg 1, 2628CJ Delft, The Netherlands}
\affiliation{\textup{\href{r.a.norte@tudelft.nl}{r.a.norte@tudelft.nl}}}

\begin{abstract}
Mechanical frequency combs are poised to bring the applications and utility of optical frequency combs into the mechanical domain. So far, their main challenge has been strict requirements on drive frequencies and power, which complicate operation. We demonstrate a straightforward mechanism to create a frequency comb consisting of mechanical overtones (integer multiples) of a single eigenfrequency, by monolithically integrating a suspended dielectric membrane with a counter-propagating optical trap. The periodic optical field modulates the dielectrophoretic force on the membrane at the overtones of a membrane's motion. These overtones share a fixed frequency and phase relation, and constitute a mechanical frequency comb. The periodic optical field also creates an optothermal parametric drive that requires no additional power or external frequency reference. This combination of effects results in an easy-to-use mechanical frequency comb platform that requires no precise alignment, no additional feedback or control electronics, and only uses a single,~\si{\milli\watt} continuous wave laser beam. This highlights the overtone frequency comb as the straightforward future for applications in sensing, metrology and quantum acoustics.
\end{abstract}

\date{\today}
\maketitle

\section{Introduction}
Over the last quarter century, optical frequency combs have become key tools for metrology, timing and spectroscopy~\cite{Cundiff2003,Fortier2019}, and are indispensable in many laboratories around the world. The fixed frequency and phase relations between the many different tones of a comb have revolutionized fields as astronomy~\cite{Metcalf2019} or cosmology~\cite{BACON2021}, and allowed tests of fundamental physics with atomic clocks~\cite{Rosenband2007,BACON2021}. Recent developments in optomechanics have expanded methods to create optical frequency combs by interactions with mechanical resonators~\cite{Miri2018,Mercade2020,Zhang2021,Wu2022,Han2022}. But in the last few years, a new paradigm of frequency combs has appeared that are completely mechanical in nature: Phononic frequency combs~\cite{Mahboob2012,Cao2014,Maksymov2022}, also called acoustic or mechanical frequency combs. The ideas behind this field began in nonlinear dynamics, where it was realized that mixing in coupled oscillators may lead to a series of sidebands~\cite{Erbe2000}, that can be regarded as a frequency comb if there exists a fixed phase relation~\cite{Cao2014} between these sidebands. Experimental demonstrations have shown mechanical frequency combs exist in different mechanical systems~\cite{Mahboob2016,Seitner2017,Ganesan2017,Czaplewski2018,Park2019,Goryachev2020,Singh2020,Chiout2021,Ochs2022,Han2022},
and have explored connections to well-known concepts in nonlinear dynamics such as bifurcations~\cite{Mahboob2016,Czaplewski2018,Batista2020}, 3- or 4-wave mixing~\cite{Erbe2000,Ganesan2017,Ganesan2017a,Ganesan2018}, and symmetry-breaking~\cite{Keskekler2022}. The fixed frequency and phase relation between the comb teeth allows the application of techniques known from optical combs. These can e.g.~improve position sensing accuracy in optically opaque materials~\cite{Wu2019} such as underwater or medical imaging. Mechanical frequency combs can further be used to track and stabilize mechanical resonances~\cite{Ganesan2019,Wall2020}, and enhance Brillouin microscopy~\cite{Aleman2020,Muralidhar2020}. Recent proposals from the field of quantum acoustics~\cite{Chu2017} foresee an important role for devices with multiple mechanical modes with equal frequency spacing. These may also be useful in scaling up transduction between optical and microwave frequencies~\cite{Bochmann2013}, and engineering the interference between subsequent modes may aid in coupling phonons to multiple qubits~\cite{Sletten2019}. Until now, mechanical frequency combs have been hamstrung by the (generally) non-linear phononic dispersion relation, which demanded high drive powers, carefully designed mode frequencies or engineered mechanical non-linearities to obtain an evenly spaced set of modes.



In a different regime of breakthrough physics, optical trapping has allowed us to manipulate and control small particles ranging from single atoms~\cite{Hu1994,Endres2016} to micrometers~\cite{Dholakia2011} in size. These particles are confined by the potential created by a strongly focused laser, which has enabled exploration of cutting-edge fields in both fundamental physics and biology. Using optical traps (tweezers), biologists can precisely manipulate anything from single strands of DNA~\cite{Wang1997} to whole living cells~\cite{Ashkin1989}. Optically trapped ultra-cold atoms and levitated nanoparticles are perfect testbeds for fundamental physics involving gravity and mesoscopic quantum mechanics~\cite{Guccione2013,Bose2017,Delic2020}. Although optical trapping and frequency combs are widespread techniques, there is little direct overlap between these regimes of physics.


In this work, we uniquely interface optical traps with frequency combs via a mechanism that enables mechanical frequency combs without requiring feedback control, external drives and frequency references, or precision optics. We observe that due to a weak counterpropagating optical trap~\cite{Ashkin1970,Zemanek1999}, strongly driven silicon nitride (Si$_3$N$_4$) membranes vibrate not only at their mechanical eigenfrequencies, but at perfect integer multiples of a single frequency, which can form a mechanical frequency comb. The standing-wave optical field exerts a dielectrophoretic force on the membrane, which is modulated by the membrane's motion as it crosses extrema of the optical field. If the displacement is small, this interaction can suppress mechanical dissipation~\cite{Ni2012}, but if the displacement becomes of the order of a quarter wavelength, it creates integer multiple copies of the original membrane motion ("overtones"~\cite{Yang2019}, see the supplementary information (SI) Sec.~A). These overtones share a fixed frequency and phase relation, and thus form a frequency comb while avoiding all difficulties of engineering a linear mechanical dispersion relation. Strikingly, the overtones are solely dependent on the amplitude of motion and the optical field, and thus independent of drive mechanism. We utilize an optothermal parametric drive based on the same single, unmodulated optical field to bring the membrane to self-oscillation. This allows us to build a mechanical comb without any additional pump tone or frequency reference, which makes overtone combs uniquely simple to generate. We will first describe the mechanism that creates the overtones and study their behavior in the frequency domain. Then, in the the time domain we will show the fixed phase relation that makes the overtones act as a frequency comb.

\section{Results}
\subsection{Overtone and driving mechanism}
\begin{figure}
\includegraphics[width = 0.5\textwidth]{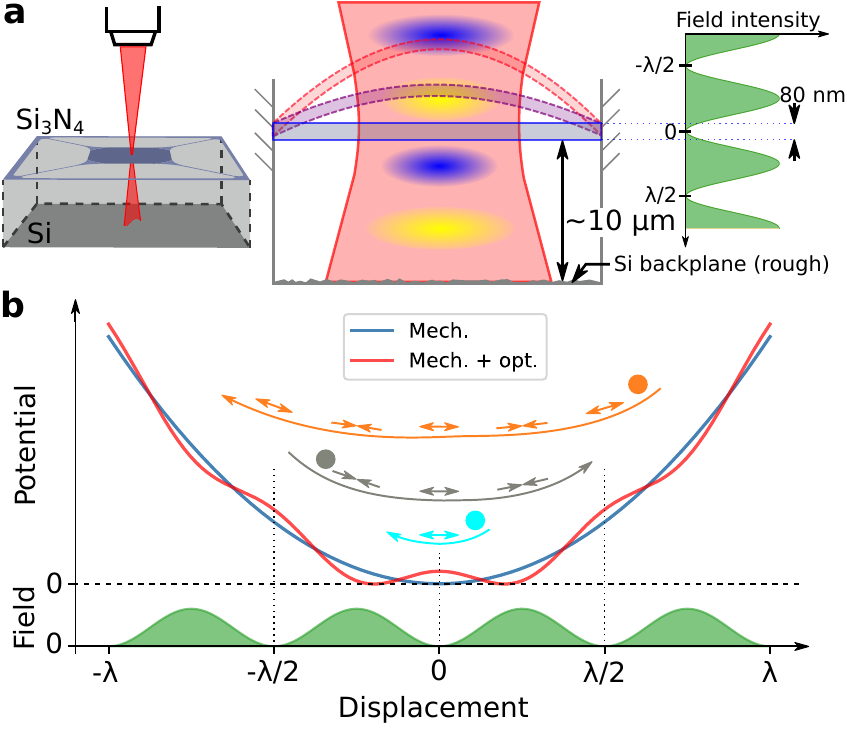}
\caption{\textbf{Schematic of overtone frequency comb. a:} The suspended Si$_3$N$_4$ membrane is subject to an out-of-plane laser, focused through a microscope objective. Part of the light is reflected from the Si backplane and interferes with the incident light, creating a counterpropagating wave optical trap such that the field intensity (green) is periodic. The membrane is clamped to the substrate, and motion is predominantly out-of-plane (purple, red dashed lines). \textbf{b}: The optical field intensity causes a (spatially) periodic modulation of the elastic potential through the dielectrophoretic force (red, blue lines). For small motion of a membrane at frequency $\omega_0$ (cyan), the restoring force component from the optical field switches sign twice per oscillation (arrow pairs), same as the elastic potential. However, for larger membrane motion (grey, orange), more optical extrema are crossed so the optical field component switches sign multiple times per oscillation, efficiently generating frequency components at $n\omega_0$ ($n=2,3,4,...$) which form an overtone frequency comb.}
\label{Schematiccomb}
\end{figure}
In this section, we first describe the physical system and the mechanism that creates the overtone frequency combs, and then introduce a more quantitative model for their dynamics. The system consists of a suspended $t=$~\SI{80}{\nano\meter} thick Si$_3$N$_4$ trampoline membrane~\cite{Norte2016} (see Methods), shown schematically in Fig.\ref{Schematiccomb}\textbf{a}. It rests $\sim 10$~\si{\micro\meter} above a backplane formed by the silicon (Si) substrate, and a laser ($\lambda = 633$~\si{\nano\meter}, $p \lesssim 3$~\si{\milli\watt}) from a commercial Polytec MSA400 laser Doppler vibrometer is incident on the membrane. At this wavelength, the Si$_3$N$_4$ reflects $\lesssim 30\%$ of the light, and Si reflects about $35\%$. Part of the light thus forms a standing wave, Fig.~\ref{Schematiccomb}\textbf{a}, periodic in the direction of the mechanical motion. 

The dielectric Si$_3$N$_4$ experiences a dielectrophoretic force proportional to the gradient of the optical intensity, similar to a particle in a counterpropagating-wave optical trap~\cite{Ashkin1970,Zemanek1999}. The trap also exerts a radiation pressure force, but this is negligible in our system (\cite{Miri2018} and SI Sec.~C). If the dielectric moves (e.g. by driving a mechanical eigenmode), it will experience a restoring force from its own elastic potential (blue solid line in Fig.~\ref{Schematiccomb}\textbf{b}), with an additional component from the optical field (red solid line). For small motion ($|x| \ll \lambda/4$, cyan line in Fig.~\ref{Schematiccomb}\textbf{b}), the optical potential functions as an additional spring~\cite{Ni2012}. However, if the motion of the membrane is of the order of the optical potential period ($\lambda/2$), the modulated potential generates the overtones.

This can be seen as follows: The restoring force of the mechanical potential switches sign twice per oscillation (arrow pairs in Fig.~\ref{Schematiccomb}\textbf{b}). If the motion is large enough (grey, orange lines) for the resonator to cross multiple extrema of the optical field, the optical component of the restoring force switches direction $2n$ times per oscillation ($n$ integer number of optical extrema). If the original motion was at mechanical eigenfrequency $\omega_0$, this effect generates motional components at $n\omega_0$, which are the overtones of the original eigenmode. We will show in Sec.~\ref{SecCombdynamics} that these overtones have a fixed phase relation and thus form a mechanical frequency comb. Because the overtones originate from the combination of the optical and mechanical potential, they completely avoid the difficulties of engineering the mechanical dispersion while still resulting in perfectly evenly spaced tones. In contrast to other combs, the tones do not exist around some carrier frequency.

\begin{figure*}
\includegraphics[width = 1.0\textwidth]{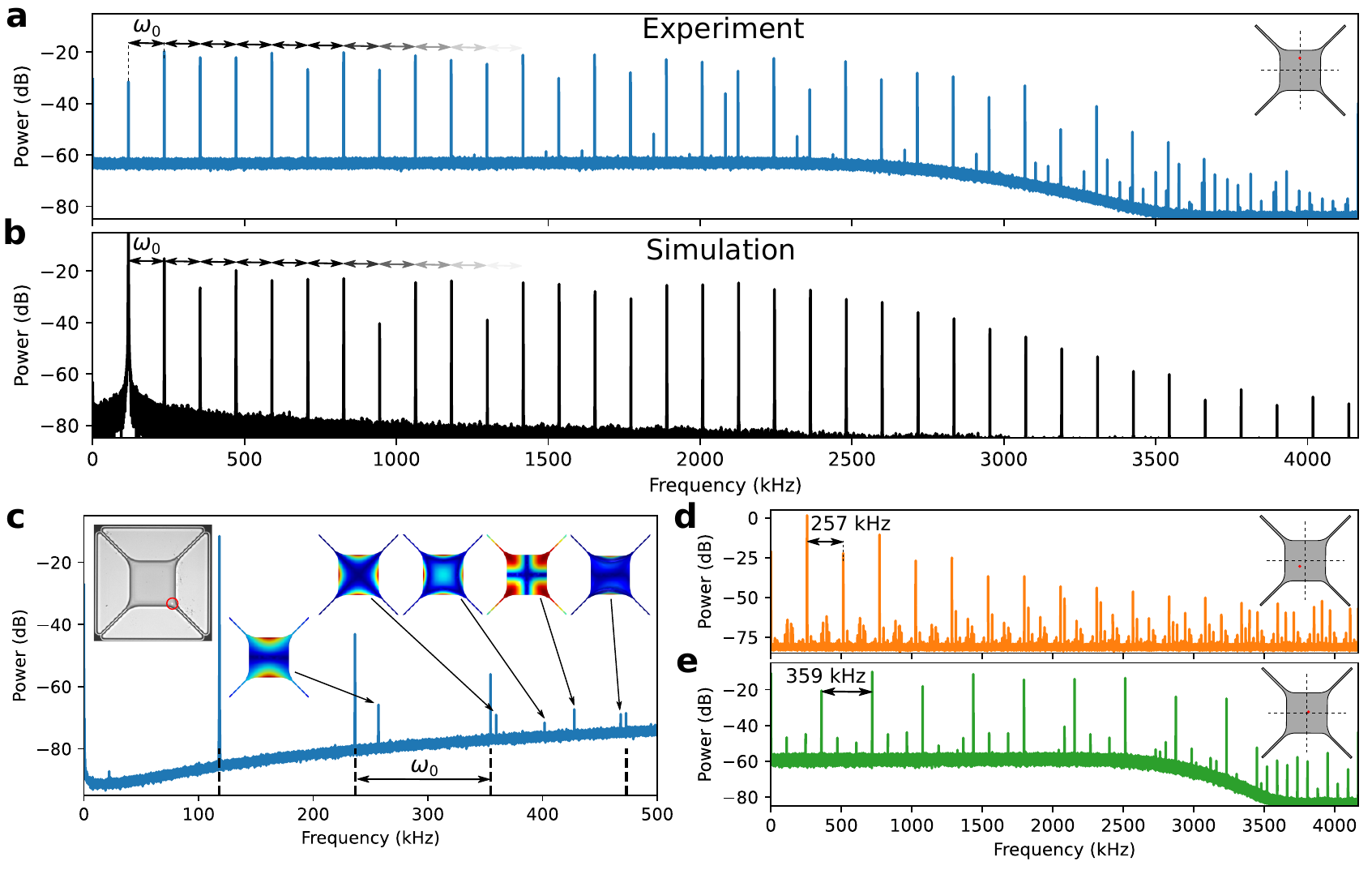}
\caption{\textbf{Overtone frequency combs. a}: Measured spectrum of a mechanical frequency comb of the fundamental mode ($\omega_0 = 2\pi\times 118.049$~\si{\kilo\hertz}) of a suspended Si$_3$N$_4$ membrane, consisting of 35 peaks spaced by $\omega_0$. The overtones are spectrally flat until $\sim 2200$~\si{\kilo\hertz}, after which their amplitude drops exponentially. Inset shows the location of the laser spot to generate and read out the comb. The additional peak around \SI{2083}{\kilo\hertz} are from a different mechanical mode, see SI Sec.~F. \textbf{b}: Simulated overtone comb spectrum, by integrating Eq.~\eqref{EOM}, matching the measurement of \textbf{a}. \textbf{c} Simultaneous measurement of fundamental-mode frequency comb (first four teeth) and higher-order mechanical modes of the membrane. Insets: Laser spot position (left) and mode shapes. \textbf{d,e} Mechanical frequency combs of the second mode (\SI{257}{\kilo\hertz}) and third modes (\SI{359}{\kilo\hertz}). Difference in the noise floor is due to different decoders used due to overloading. The \SI{257}{\kilo\hertz} mode has a near-degeneracy, so the comb spectrum is less clean. Insets show the laser position to drive and read out the different modes.}
\label{Frequencycomb}
\end{figure*}

We model the overtone frequency comb using a resonator described by displacement $x(t)$ ($x =0$ mechanical equilibrium), with resonance frequency $\omega_0$ and decay rate $\gamma$, where we have divided by the simulated effective mass $m_\mathrm{eff} \simeq 12\times10^{-12}$~\si{\kilo\gram}. We write the dielectrophoretic force as a proportionality constant $F_\mathrm{o}$ (units of force) times the gradient of the periodic part of the optical intensity, $\nabla E^2 \propto \sin\left(\frac{4\pi}{\lambda}(x-x_\mathrm{off})\right)$. This way we can move finite-size effects of the membrane into $F_\mathrm{o}$, which we numerically evaluate in SI Sec.~C. We obtain the equation of motion
\begin{equation}
\ddot{x} + \gamma \dot{x} + \omega_0^2 x = \frac{F_\mathrm{o}}{m_\mathrm{eff}} \sin\left(\frac{4\pi}{\lambda}(x-x_\mathrm{off})\right),
\label{EOM}
\end{equation}
which requires only a suitable initial condition $|x|_{t=0} \gtrsim \lambda/4$ to demonstrate the creation of overtones. This condition is much larger than typical interferometric position measurements, which is why we use a laser Doppler vibrometer that is capable of resolving such large displacements (see Methods). Eq~\eqref{EOM} also shows that the overtones are independent of the choice of drive (e.g.~piezoelectric, electrostatic, thermal). By utilizing the spatially-periodic optical field through optothermal effects~\cite{Aubin2004}, we can create a parametric drive powerful enough to bring the membrane to self-oscillation. That is, when the resonator moves through the field, the optical intensity it experiences is modulated at twice the frequency of the original motion. This modulates the resonator frequency through absorption (thermal expansion), thus creating an optothermal parametric drive (see SI Sec.~D). Our membranes are patterned with a specific photonic crystal that enhances the absorption of \SI{633}{\nano\meter} laser light. This allows a single, continuous-wave laser beam of \si{\milli\watt} power to bring the membrane into self-oscillation. While we verify that the overtones can also be generated via inertial (piezo) driving (see SI Sec.~E), we leverage the optothermal drive to avoid any external pump tone or frequency reference. This property of overtone combs is unique within the mechanical frequency combs, and allows for significantly simpler setups.

\subsection{Overtone frequency comb}
We measure the overtone frequency comb for the fundamental mode $\omega_0 = 2\pi \times 118.049$~\si{\kilo\hertz} of our membrane in Fig.~\ref{Frequencycomb}\textbf{a}. In the velocity power spectrum, we see a series of 35 peaks spanning the full detection bandwidth of the setup (\SI{4.2}{\mega\hertz}), spaced by $\omega_0$. This spacing is perfectly uniform, limited by the spectral resolution of the setup to $4.7\times 10^{-8}$ relative spacing difference. A comparison of this comb to other mechanical combs is included in the SI Sec.~A. Through integration of the velocity signal, we verify the displacement $x \gg \lambda/4$. By numerically integrating Eq.~\eqref{EOM}, we can reproduce this overtone comb, shown in Fig.~\ref{Frequencycomb}\textbf{b}, using only the optical potential strength $F_\mathrm{o}$, offset $x_\mathrm{off}$ and initial conditions $[x_0, v_0]$ as fit parameters (assuming steady state, so $\gamma = F_\mathrm{d} = 0$). This highlights that the non-linearity comes from the optical field, without introducing mechanical nonlinearities previously used to explain this behavior~\cite{Yang2019}. 
\begin{figure*}
\includegraphics[width = 1.0\textwidth]{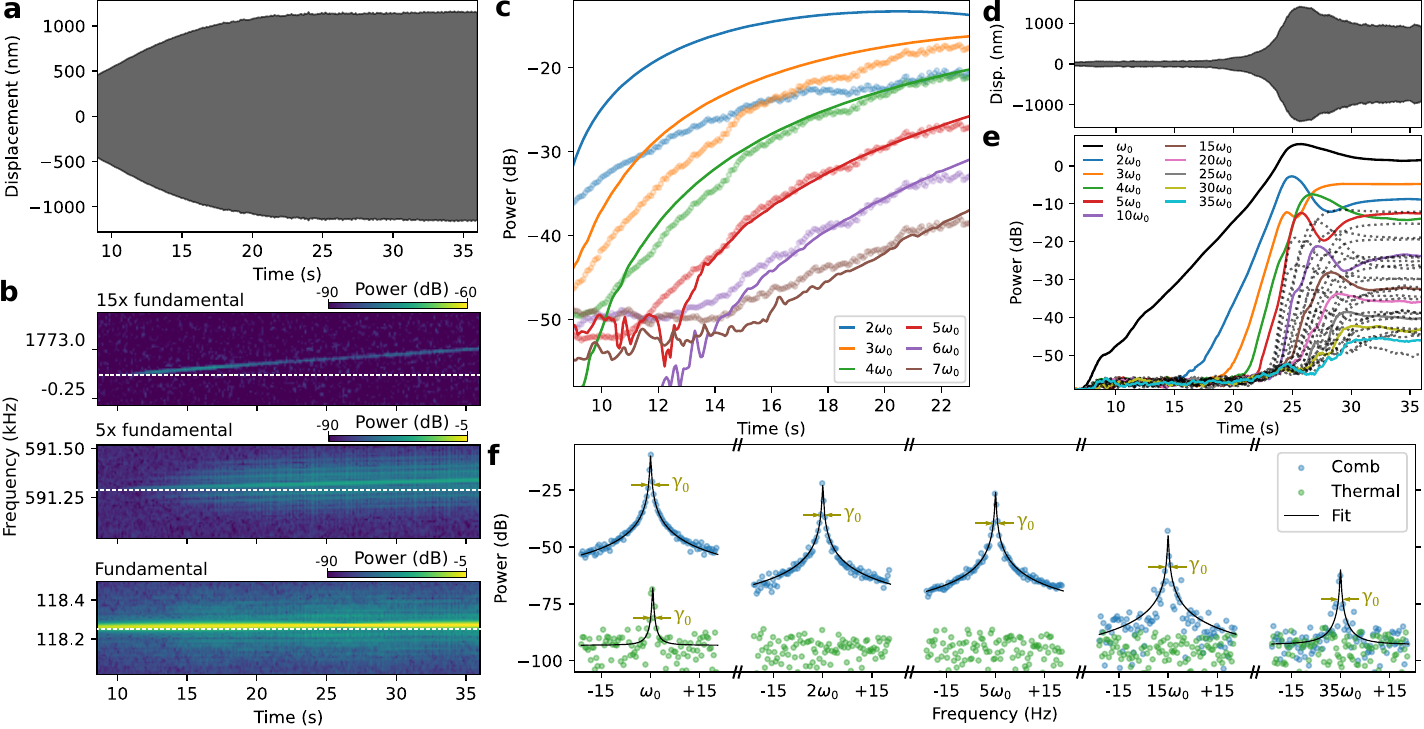}
\caption{\textbf{Comb dynamics. a}: Measured displacement of membrane, showing clear growth before reaching a plateau around $t = 24$~\si{s}. \textbf{b}: Measured spectrum of membrane motion close to fundamental mode and two of its overtones (5th and 15th). A thermal shift of \SI{9}{\hertz} of the fundamental mode is visible as a $5\times9$ and $15\times9$~\si{\hertz} shift in the overtones (white dashed lines are horizontal to guide the eye). \textbf{c}: First six overtones increasing in power as the membrane displacement increases (markers) in \textbf{a}, with simulated dynamics (solid lines) based on integration of Eq.~\eqref{EOM} matching quantitatively to the highest four. \textbf{d}: Membrane displacement of a different device (different chip) than \textbf{a}, showing growth from close to thermal regime to steady state. \textbf{e}: Overtone amplitudes extracted from the spectrum of measurement of \textbf{d}, showing all 35 overtones within the detection bandwidth (black, dotted lines are overtones with numbers between the labeled ones).  \textbf{f}: Spectrum showing measured fundamental mode and some selected overtones in the comb regime (blue) and thermal (green), along with Lorentzian fit with center frequencies $n\omega_0$ ($n$ integer) and identical linewidths $\gamma_0$.}
\label{Slowshift}
\end{figure*}

The overtones can be distinguished from the other mechanical eigenmodes of the membrane. The mechanical eigenmodes can be detected and identified (Fig.~\ref{Frequencycomb}\textbf{c}). When the overtone comb is generated, additional peaks appear in the spectrum (dotted black lines). This demonstrates that the overtones are not affected by the mechanical dispersion relation and do not require engineered non-linear resonances~\cite{Cao2014,Ganesan2017}. Furthermore, it is possible to generate combs from the second and third mechanical eigenmodes, by selecting the right laser position on the membrane, Fig.~\ref{Frequencycomb}\textbf{d,e}. This makes the frequency spacing variable, limited by our ability to drive a particular eigenmode. Finally, for some laser positions the overtone comb at $\omega_0$ interacts with a different mechanical eigenmode ($\omega_\mathrm{h}$), which could allow extension of the comb bandwidth (SI Sec.~F).

We examine Eq.~\eqref{EOM} to better understand the behavior of this mechanism (details in SI Sec.~A). Firstly, when the displacement $x$ increases, more overtones appear due to the increasing number of optical extrema crossed. Each overtone will again be modulated, so they have equal power up to a certain cutoff, which is beneficial for many applications ($\sim 2200$~\si{\kilo\hertz} in Fig.~\ref{Frequencycomb}\textbf{a,b}). Secondly, the strength of the optical field ($F_\mathrm{o}$) controls the power of each overtone relative to the original mode $\omega_0$. Finally, the offset $x_\mathrm{off}$ between mechanical and optical zeros determines the relative intensity of the odd and even number overtones. The position offset $x_\mathrm{off}\simeq 40$~\si{\nano\meter} is consistent between different membranes in this work, and is half the thickness of the Si$_3$N$_4$ membranes. We can repeatably create the overtone comb in different membranes, and study the effect of the optical beam itself on the membranes and overtone comb in SI Sec.~G.  

\subsection{Comb dynamics}\label{SecCombdynamics}
The comb dynamics can be visualized by starting a measurement with the laser spot positioned away from the membrane, as shown in Fig.~\ref{Slowshift}\textbf{a}. We move the laser to the membrane center at $t=$~\SI{8.7}{\second} such that the optothermal parametric driving starts increasing displacement. We analyze the dynamics by cutting the recorded time signal into intervals and performing a Fourier transform on each. Then we concatenate the spectra such that we can study its behavior over time (Fig.~\ref{Slowshift}\textbf{b}) and monitor the power of individual overtones by taking linecuts (Fig.~\ref{Slowshift}\textbf{c}). 

In Fig.~\ref{Slowshift}\textbf{b}, the fundamental mode shows a \SI{9}{\hertz} upwards shift in frequency, which is reproduced as an $n\times9$~\si{\hertz} upwards shift for the $n$th overtone (e.g. $n=5,15$ in the figure) and likely originates from slow thermalization (SI Sec.~D). This illustrates thermal tuning would be an effective mechanism to control and tune the comb spacing, without compromising comb uniformity. 

We plot the power of the overtones as the comb grows in Fig.~\ref{Slowshift}\textbf{c}. We can simulate and reproduce this growth quantitatively by integrating Eq.~\eqref{EOM}. Comparison between the experimental data (markers) and simulation (solid lines) in Fig.~\ref{Slowshift}\textbf{c} shows good agreement with fit parameters [$F_\mathrm{o} = 3.8$~\si{\pico\newton}, $x_\mathrm{off} = 40$~\si{\nano\meter}, $x_0 = 5$~\si{\nano\meter}, $v_0 = 1$~\si{\nano\meter\per\second}, $\gamma/2\pi = 0.8$~\si{\hertz} and $F_\mathrm{d} = 2.1$~\si{\pico\newton}]. Most important for the overtones is $F_\mathrm{o}$, which is in excellent agreement with simulated force on the order of \si{\pico\newton} (SI Sec.~C). All extracted curves share a vertical offset to account for the total detection efficiency. This shows Eq.~\eqref{EOM} reproduces the dynamics of the resonator and individual overtones.

In Figs.~\ref{Slowshift}\textbf{d,e}, we observe the membrane motion from thermal state at $t<5$~\si{\second} to the steady state of the comb at $t>33$~\si{\second}. Initially, we detect only the fundamental mode $\omega_0$, until the resonator starts crossing multiple optical extrema. Higher overtones appear as the maximum displacement grows. In contrast to Fig.~\ref{Frequencycomb}\textbf{a}, the fundamental mode is the most powerful. The detection efficiency likely varies for each overtone due to their shape~\cite{Yang2019}. The displacement stops growing at $t\simeq 26$~\si{\second}, and decreases slightly before the system reaches steady state (see SI Sec.~H). It is likely limited in amplitude by a mechanical non-linearity, but the amplitude overshoot and the oscillation of amplitude of overtones suggests that the interaction with the optical field play a role. In the steady state, the fractional frequency stability of the 30th overtone is $7.5\cdot 10^{-10}$ over a 6-hour period (SI Sec.~H), limited by thermal drifts. The frequency of the overtones determined solely by the mechanical frequency, and is thus not affected by drifts in the laser frequency. Changes in the laser power will affect the overtone amplitudes, mainly via the optothermal parametric driving. The uniformity of the comb is not affected by mechanical frequency shifts (nor by the laser), thus uniformity is constant.

We isolate and plot the spectrum around several of the overtones in Fig.~\ref{Slowshift}\textbf{f}. In the thermal regime (green markers), only the fundamental mode is visible and we can fit a Lorentzian with linewidth $\gamma_0 = 2\pi\times 0.07$~\si{\hertz} to the peak. For the comb in steady state, the fundamental mode retains its linewidth $\gamma_0$, and we can derive that the comb mechanism does not cause additional noise, SI Sec.~I. All overtones possess the same linewidth $\gamma_0$, which does not match the $n\gamma_0$ scaling expected from a frequency-fluctuation limited linewidth, but matches a decay-rate limited system. These properties combined should allow frequency combs with single-\si{\milli\hertz} linewidths based on ultra-high-Q membrane resonators~\cite{Hoj2021}. 

\begin{figure}
\includegraphics[width = 0.5\textwidth]{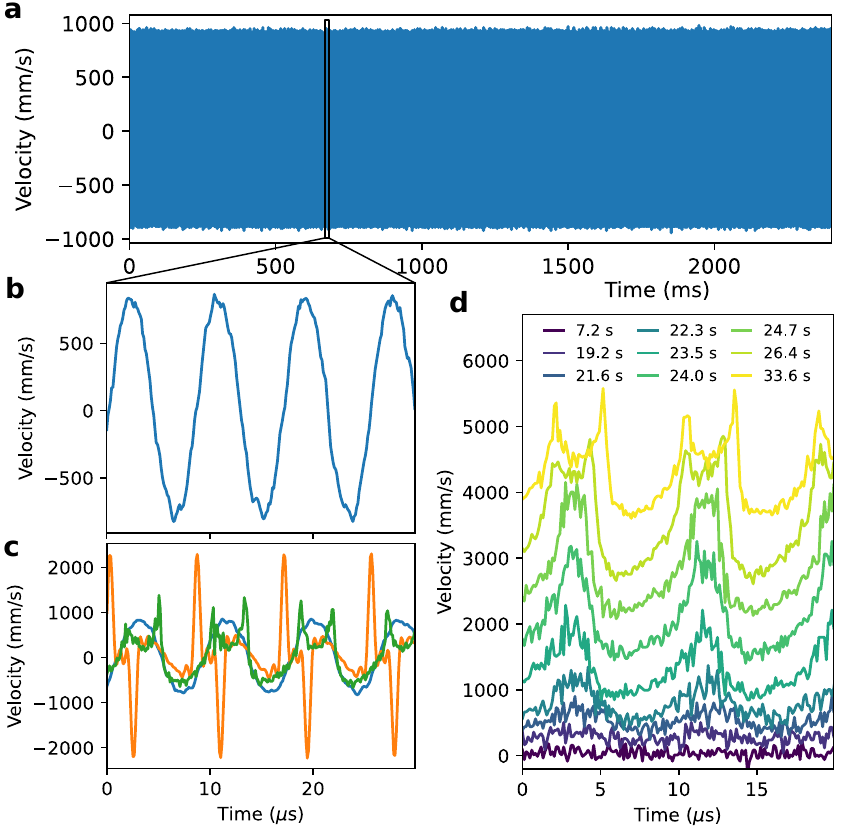}
\caption{\textbf{Time domain frequency comb. a}: Measured velocity of resonator in frequency comb regime, displaying no pattern with period longer than $1/\omega_0$. \textbf{b}: Zoom-in of time-domain signal in \textbf{a}, showing the \SI{118}{\kilo\hertz} fundamental mode dominates. \textbf{c}: Three different measured time signals in steady state, with varying relative strengths of the overtones (blue: $\omega_0$ dominates, orange: $2\omega_0$ and $3 \omega_0$ dominate, green: $1\omega_0$ to $6\omega_0$ similar in strength). All traces can be reproduced only with phase-coherent addition of the different overtones. \textbf{d} Time domain signal of comb during growth, same measurement as Fig.~\ref{Slowshift}\textbf{e} at the indicated times. This shows the smooth transition from thermal regime (bottom) to the comb in steady state (top), traces are offset vertically for clarity.}
\label{Timedomaincomb}
\end{figure}

To finally show that the overtones form a comb, i.e.~that a fixed phase and frequency relation exists, we show the time-domain signal in Fig.~\ref{Timedomaincomb}\textbf{a}. Unlike combs centered around a carrier (such as soliton-based mechanical~\cite{Ganesan2017} and optomechanical~\cite{Zhang2021} frequency combs), there is no component with periodicity longer than the mechanical period $2\pi/\omega_0$ (Figs.~\ref{Timedomaincomb}\textbf{b,c}). This highlights the different physical and dynamical processes behind the overtone comb. If the fundamental mode is the dominant component in the comb, we get a sinusoid (Figs.~\ref{Timedomaincomb}\textbf{b,c}, blue line). When other components are dominant, particularly $2\omega_0$ and $3\omega_0$, we get the peaked curves (Fig.~\ref{Timedomaincomb}\textbf{c}, green and orange lines), which retain the periodicity of the fundamental mode. The shape of these curves proves that the entire overtone comb is phase-coherent, as they are formed by a sum of cosines with the same phase offset (see SI Sec.~J). This behavior is retained not only in the steady-state, but also during comb growth, Fig.~\ref{Timedomaincomb}\textbf{d}. There we plot the time signal at various stages during the measurement of Fig.~\ref{Slowshift}\textbf{d,e}, which shows a smooth transition from sinusoidal to peaked behavior as the comb grows. Thus the overtones form a mechanical frequency comb.

\section{Discussion}
We have discovered a mechanism which uniquely interfaces two breakthrough concepts: optical trapping and frequency combs. This allows for mechanical frequency combs whose simplicity stands out from those based on previous mechanisms. This is realized by integrating a suspended dielectric membrane in a weak optical trap. The dielectrophoretic force from the optical field modulates the mechanical potential of the membrane. This modulation creates integer multiple copies (overtones) of the membrane's motion, which forms a frequency comb. We show combs of up to 35 overtones in a \SI{4.2}{\mega\hertz} bandwidth with control over the frequency spacing, excellent uniformity, stability and no added mechanical noise. The integration of the membrane in the optical trap allows us to combine the overtone comb with optothermal parametric driving, which brings the membrane to self-oscillate. We thus realize a frequency comb that requires no external drive or control frequencies and which uses a minimal setup (laser, microscope objective and vacuum chamber). This makes it more versatile and easier to use than other ways of generating these combs. In summary, this mechanism unlocks potential of mechanical frequency combs for sensing, timing and metrology applications at the microscale, and provides native integration with phononic circuits.  


\section{Methods}\label{SecMethods}
\subsection{Laser Doppler Vibrometer}
\begin{figure}
\includegraphics[width = 0.5\textwidth]{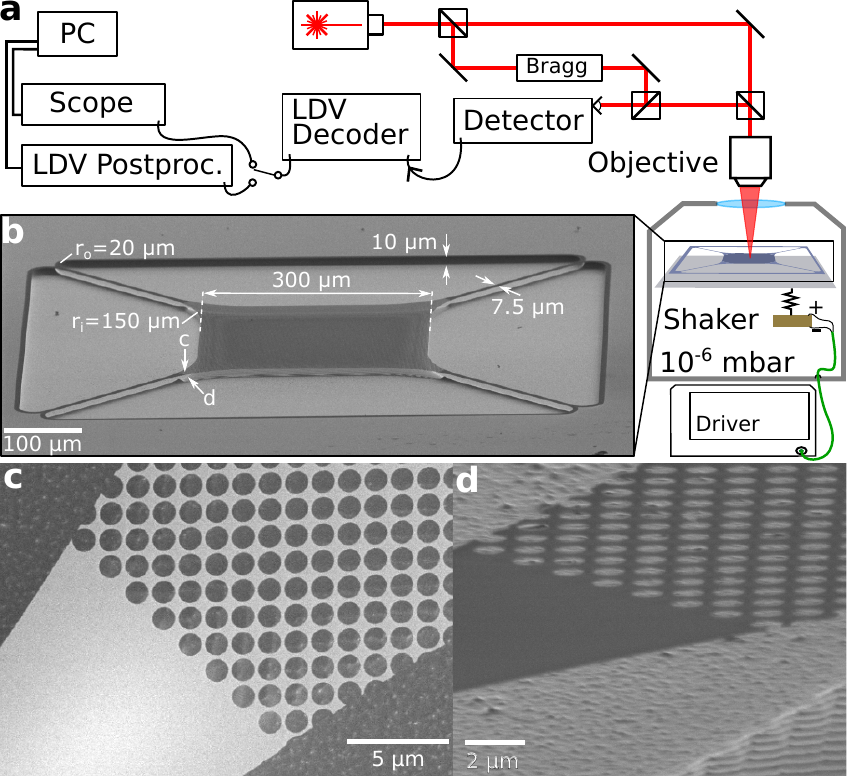}
\caption{\textbf{Setup and membranes. a}: Schematic of the laser Doppler vibrometer (LDV) setup, where we access the chip containing the membranes using a microscope objective directly outside a vacuum chamber. \textbf{b}: SEM image of the suspended membrane, with nominal design parameters in white. \textbf{c}: Top-down view of corner of the membrane, highlighting the interface of the photonic crystal with the tether. \textbf{d}: Tilted view to show Si surface below \textbf{c}, to illustrate etch roughness and an etched imprint of the photonic crystal pattern (bottom right).}
\label{Setupandmembrane}
\end{figure}
The setup used in this work is shown in Fig.~\ref{Setupandmembrane}\textbf{a}. It consists of a commercial laser Doppler vibrometer (LDV), Polytec MSA400, which is depicted schematically. Light from the LDV goes through a microscope objective, which focuses it on the chip containing the membrane resonators. This chip is placed in a vacuum chamber and pumped down until the pressure is $<5\times 10^{-6}$~\si{\milli\bar}, to reduce gas damping. There is a piezoelectric shaker mounted to the sample holder, by which we can drive the membrane, though for the majority of the measurements we use the thermal parametric driving described in the SI Sec.~D. The reflected light from the membranes is Doppler-shifted due to their out-of-plane motion, which is then detected by the LDV. To extend the time we can continuously measure, we add a Rohde \& Schwarz RTB2004 digital oscilloscope to readout the LDV decoder. Using the history function of this oscilloscope, we can chain 16 measurements of 20 million data points each, which allows \SI{38}{\second} of time signal at \SI{8.33}{\mega\hertz} sampling rate. However, this method for reading out the velocity comes at a cost of the calibrated readout that the LDV postprocessing offers.

The LDV outputs a voltage signal proportional to velocity or displacement, depending on the LDV decoder used. If using the LDV postprocessing, it can easily be Fourier-transformed and exported. If using the oscilloscope and history function, we sequentially read out the measurements afterwards and concatenate them in the correct order in post-processing. We can then further extract information from this signal either by integrating (to obtain the displacement), or by using Scipy's short-time Fourier transform function to obtain the time-varying behavior of the comb spectrum. 

\subsection{Trampoline membranes}
The membrane structures used in this work are fabricated out of \SI{100}{\nano\meter} thick stoichiometric low-pressure chemical vapor deposition Si$_3$N$_4$ on top of a \SI{1}{\milli\meter} Si chip. This was done by writing the pattern using electron beam lithography and an inductively coupled plasma (ICP) etch to transfer that pattern to the Si$_3$N$_4$. The membranes are then released using a second ICP etch, at \SI{-120}{\celsius} for \SI{30}{\second}, resulting in about \SI{10}{\micro\meter} of undercut and a final Si$_3$N$_4$ thickness of \SI{80}{\nano\meter}. A SEM image of a released trampoline is shown in Fig.~\ref{Setupandmembrane}\textbf{b}, with the nominal design parameters added in white. The resulting trampolines have well-characterized mechanical properties~\cite{Norte2016,deJong2022b}, with fundamental mode frequencies at \SI{120}{\kilo\hertz} and Q-factors typically of 1 million. 

The membranes are pattered with a periodic array of holes (Fig.~\ref{Setupandmembrane}\textbf{c}) that forms a photonic crystal, though it is not designed to have a high reflectivity at the operating wavelength of our LDV ($\lambda = 633$~\si{\nano\meter}). Instead, the holes function as etch release holes to evenly release the membrane in the final ICP etch, which is essential for the fabrication yield. For these membranes, the periodic array of holes functions as a method to control the mass and resonance frequencies~\cite{deJong2022b}, and possesses internal optical resonances that facilitate absorption which we detail in SI Sec.~D. Finally, the release etch imprints the photonic crystal pattern on the Si backplane, which roughens the surface. Fig.~\ref{Setupandmembrane}\textbf{d} shows the final Si surface roughness from the release etch (top left half) and the imprint below the periodic hole pattern (bottom right half).

\section*{Data Availability}
The data (raw data, analysis and calculation scripts, and finite element simulations) that support the findings of this study are available at \href{https://doi.org/10.4121/19821016.v1}{https://doi.org/10.4121/19821016.v1}.

\section*{Acknowledgements}
We would like to acknowledge Farbod Alijani, Peter Steeneken, Lior Michaeli, and Albert Schliesser for interesting discussions, Dongil Shin for help with the thermal simulations, and Wouter Westerveld and Gerard Verbiest for their comments on the manuscript. R.N. would like to acknowledge support from the Limitless Space Institute's I2 Grant. M.J., A.C., S.G., and R.N. acknowledge valuable support from the Kavli Nanolab Delft and from the Technical Support Staff at PME, 3mE Delft, in particular from Gideon Emmaneel and Patrick van Holst. 

\section*{Author contributions}
M.J. performed the experiments, simulations and data analysis. A.G. and M.J. developed the theory. A.C. developed the fabrication process and fabricated the devices. M.J., A.G., S.G., and R.N. developed the interpretation of the results. R.N. supervised the project. All authors contributed to writing and editing the manuscript.

\section*{Competing interests}
The authors declare no competing interests.

\clearpage

\setcounter{figure}{0}
\renewcommand{\thefigure}{S\arabic{figure}}
\setcounter{equation}{0}
\renewcommand{\theequation}{S\arabic{equation}}
\setcounter{table}{0}
\renewcommand{\thetable}{S\arabic{table}}

\section{Supplementary information}
This document contains the supplementary information.

\subsection{Mechanical overtones}\label{Overtones}
\begin{figure*}[ht!]
\includegraphics[width = 1.0\textwidth]{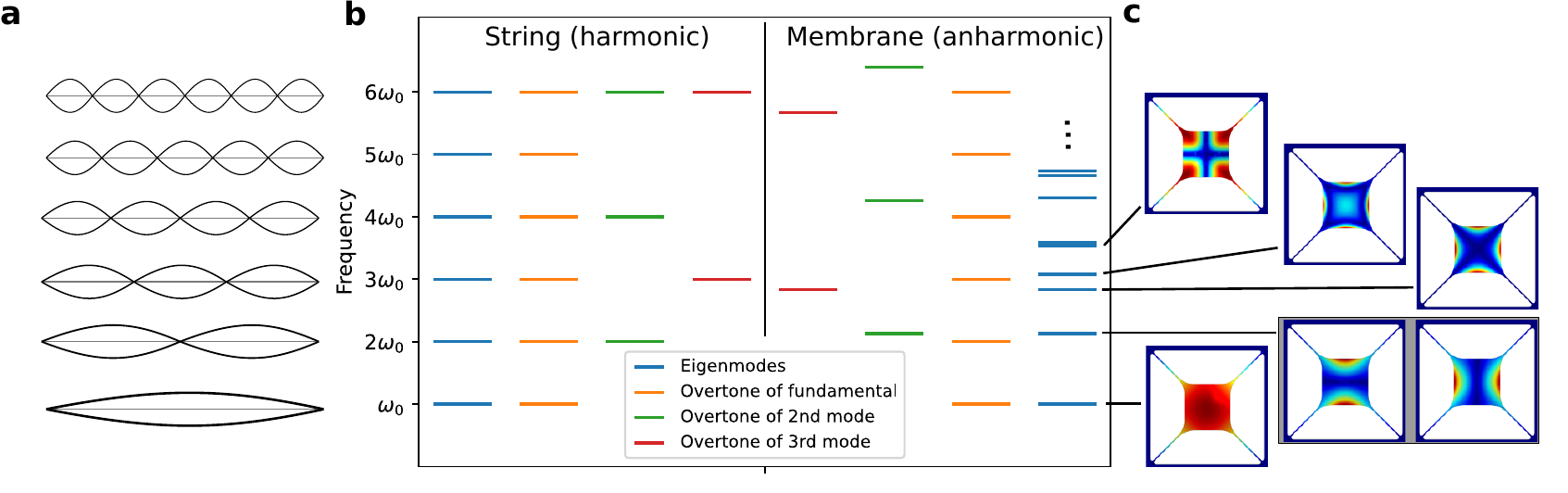}
\caption{\textbf{Mechanical overtones. a}: Resonator modes of a string. \textbf{b}: Frequency diagram of eigenmodes of a harmonic string and the anharmonic SiN trampoline membranes used in this work. \textbf{c}: SiN trampoline modes, grey box denotes degenerate modes. The color scale denotes normalized absolute displacement, from minimum (blue) to maximum (red).}
\label{Schematicovertones}
\end{figure*}
The motion of a mechanical resonator can typically be described by considering a linear superposition of the normal modes of the structure. In some structures, the normal modes are related by integer multiples of some frequency, and they are said to be \emph{harmonic}. This is the case for a model string (Fig.~\ref{Schematicovertones}\textbf{a,b}) and a desirable property for many musical instruments. However for most of the resonators in this work, the normal modes are not related by integer multiples, thus the spectrum is \emph{anharmonic}. In Fig.~\ref{Schematicovertones}\textbf{c}, we have simulated and plotted the first six eigenmodes of the trampoline membrane resonator of the main text, and plotted their frequencies in Fig.~\ref{Schematicovertones}\textbf{b}. 

In this work, we introduce an optics-based non-linearity to the equation of motion, which affects the description of the resonator motion in terms of the normal modes given by its structure. Briefly summarized, we consider only the lowest frequency normal mode given by the mechanical structure, as a simple harmonic oscillator,
\begin{equation}
\ddot{x} + \gamma \dot{x} + \omega_0^2 x = 0.
\end{equation}
The solutions for this equation are well-known, $x \propto e^{i\omega_0 t}$, which oscillates at the frequency $\omega_0$ of the normal mode. By introducing the gradient force term, $\propto \sin(x)$ (non-linear), we get terms in our solution that oscillate at integer multiples of $\omega_0$, without being related to the (other) normal modes of the mechanical resonator. To distinguish them from the other normal modes, we will refer to the components at $n\omega_0$ ($n$ integer) as \textbf{overtones}. Others have detected these components of the mechanical motion before, and refer to them in a similar manner~\cite{Zhou2017,Yang2019,Yang2021}. They ascribe the origin of these overtones to be mechanical in nature, while we propose an optical origin.

\subsubsection{Small displacement}
To expand the description of these overtones, we consider the equation of motion as Eq.~1 in the main text, where we have added a resonant drive term $F_\mathrm{d}$ and absorbed the effective mass $m_\mathrm{eff}$ into $F_\mathrm{o}$ and $F_\mathrm{d}$ for convenience. The shorthand $\Lambda = 4\pi/\lambda$ is similarly used. We get
\begin{equation}
\ddot{x} + \gamma \dot{x} + \omega_0^2 x = F_\mathrm{o} \sin\left(\Lambda(x-x_\mathrm{off})\right) + F_\mathrm{d} e^{i\omega_0 t}.
\label{SIEOM}
\end{equation}
The periodic part of the optical potential is of the form $E \propto \sin\left(\frac{2\pi}{\lambda}(x-x_\mathrm{off})\right)$. The gradient of the optical intensity, $\nabla E^2 \propto \sin\left(\Lambda(x-x_\mathrm{off})\right)$ after some algebra and absorbing the relevant constants into $F_\mathrm{o}$. The constant offset between the potentials, $x_\mathrm{off}$, can be taken out of the sine by 
\begin{equation}
\sin\left(\Lambda(x-x_\mathrm{off})\right) = \sin \left( \Lambda x\right) s_x + \cos \left(\Lambda x \right)  c_x,
\end{equation}
with $s_x = \cos(\Lambda x_\mathrm{off})$ and $c_x = \sin(\Lambda x_\mathrm{off})$.  For a small displacement $x$, we can use the Taylor series and truncate the higher order terms, so $s_x \sin(x) + c_x \cos(x) \simeq c_x + s_x x - \frac{c_x}{2} x^2$. The equation of motion then becomes
\begin{equation}
\ddot{x} + \gamma \dot{x} + \omega_0^2 x = F_\mathrm{o} \left(c_x + s_x \Lambda x - \frac{c_x}{2}\Lambda^2 x^2 \right) + F_\mathrm{d} e^{i\omega_0 t}.
\label{SIsmallamplitudeEOM}
\end{equation}
This equation admits solutions of the form 
\begin{equation}
x = \sum_{n=0}^\infty A_n e^{in\omega_0 t},
\label{Solutionforx}
\end{equation}
which are the integer multiples of our original frequency $\omega_0$; the overtones. The term $n = 0$ corresponds to a static position offset from the zero of the mechanical potential. Substituting the solution Eq.~\eqref{Solutionforx} into Eq.~\eqref{SIsmallamplitudeEOM} and gathering all terms by their frequency ($e^{in\omega_0 t}$ for every $n$ separately) allows us to extract the amplitudes of the individual overtones in terms of the parameters of our system. We get
\begin{equation}
\begin{aligned}
A_0 &\simeq c_x F_\mathrm{o} / \omega_0^2 \\
A_1 &= \frac{F_\mathrm{d}}{i\gamma \omega_0 - \Lambda F_\mathrm{o} s_x + F_\mathrm{o} \frac{c_x}{2} \Lambda^2 A_0}, \\
A_2 &= \frac{F_\mathrm{o}\frac{c_x}{2}\Lambda^2 A_1^2}{3\omega_0^2 - 2i\gamma \omega_0 + F_\mathrm{o} s_x \Lambda - F_\mathrm{o} \frac{c_x}{2} \Lambda^2 A_0}, \\
A_n &= \frac{F_\mathrm{o} \Lambda \sum_{j,k = 0}^\infty A_j A_k}{(n^2-1)\omega_0^2 - i\gamma n\omega_0 + F_\mathrm{o} s_x \Lambda - F_\mathrm{o} \frac{c_x}{2} \Lambda^2 A_0}, 
\end{aligned}
\label{SIAforsmallamp}
\end{equation} 
where the summation contains only the terms where $j+k = n$. This sequence of overtone amplitudes is monotonically decreasing (see Fig.~\ref{Overtoneamplitudes}): Every $A_n <  A_{n-1}$. In the small-displacement case, the overtones will be negligible compared to $A_1$.

\subsubsection{Large displacement}
If instead the displacement is not small, we can still use the same method. We need to keep all the terms of the Taylor expansion,
\begin{equation}
\sin(x) = \sum_{n=0}^\infty \frac{(-1)^n x^{2n+1}}{(2n+1)!}, \quad \cos(x) = \sum_{n=0}^\infty \frac{(-1)^{n} x^{2n}}{(2n)!}.
\end{equation}
We can use the same ansatz of Eq.~\eqref{Solutionforx}, and extract the amplitudes by collecting all terms of the same frequency. For $n = 0$, the solution converges to 
\begin{equation}
A_0 \simeq c_x F_\mathrm{o} /\omega_0^2,
\end{equation}
as the contribution from the higher order terms of the expansion scale with $1/\Lambda^n$. Similarly, the denominator for $n > 0$ can be truncated to obtain
\begin{equation}
\begin{aligned}
A_1 &= \frac{F_\mathrm{d}}{i\gamma\omega_0 - \Lambda F_\mathrm{o}s_x + F_\mathrm{o} \frac{c_x}{2}\Lambda^2 A_0} \\
A_n &= \frac{F_\mathrm{o} \sum_j \left(s_x s_j \sum\limits_{k,\ell,... = 0}^\infty A_{k,\ell,...}^{j} + c_x c_j\sum\limits_{k,\ell,... = 0}^\infty A_{k,\ell,...}^{j}\right)}
{(n^2-1)\omega_0^2 - i\gamma n\omega_0 + \Lambda F_\mathrm{o}s_x - F_\mathrm{o} \frac{c_x}{2}\Lambda^2 A_0},
\label{SIAforbigamp}
\end{aligned}
\end{equation}
with $s_j = \frac{(-1)^{(j-1)/2} \Lambda^{j}}{j!}$ for odd $j$ ($s_j = 0$ for even $j$) and $c_j = \frac{(-1)^{j/2} \Lambda^{j}}{j!}$ for even $j$ ($c_j = 0$ for odd $j$). The summation only contains the terms of $j$ amplitudes ($A_k, A_\ell, ...$) whose indices $k,\ell,...$ add up to $n$. To clarify, for $n = 3$, we sum over $A_1 A_1 A_1$, $A_1 A_2$, $A_2 A_1$, $A_3 A_0$, $A_0 A_3$ as well as many terms with more $A_0$'s. For every $n$, we get an infinite number of contributions to each term. The denominator contains at most two terms that depend on $n$, but the rest is constant and the whole denominator converges for each $n$; thus we can truncate the denominator, as we have done in Eq.~\eqref{SIAforbigamp}. In contrast, to evaluate the numerator we can take sequentially higher terms in the Taylor expansion (i.e.~increase $j$), find all combinations of $j$ integers that sum to $n$, and add them to the term of frequency $n\omega_0$.

\begin{figure}
\includegraphics[width = 0.5\textwidth]{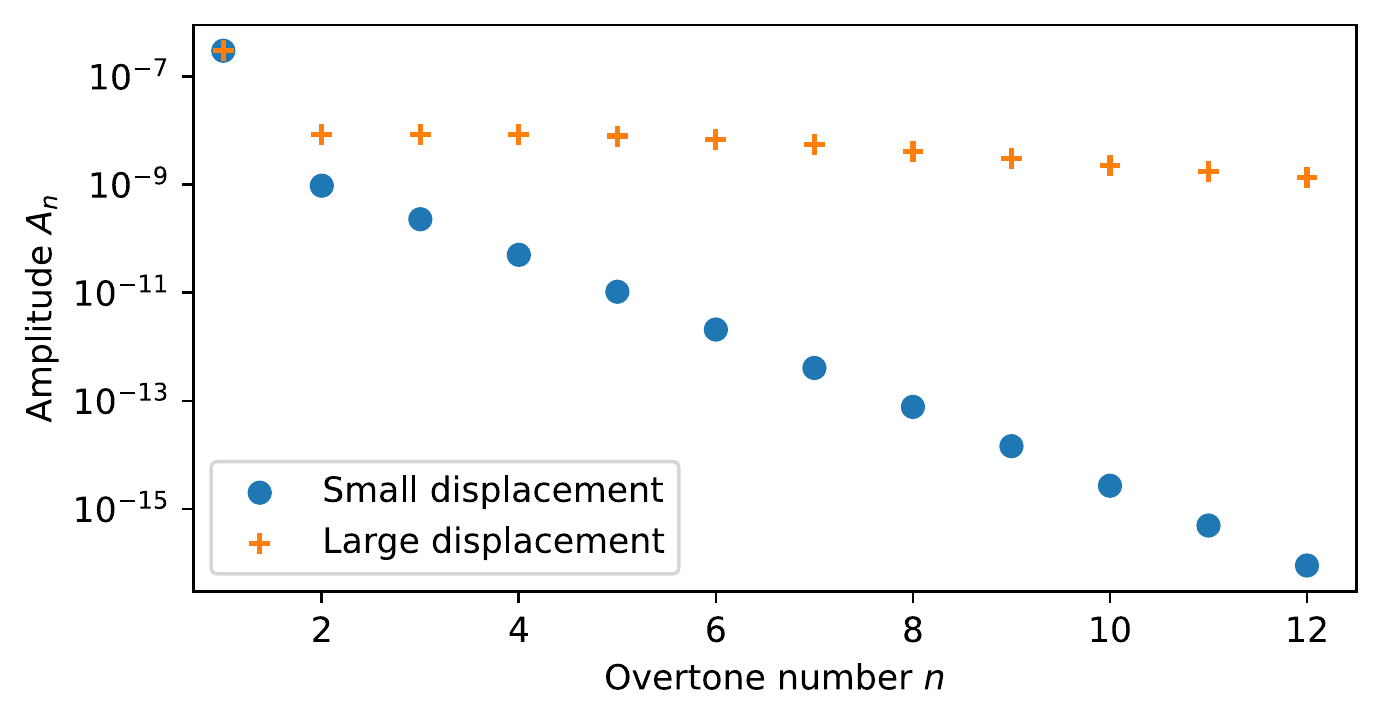}
\caption{\textbf{Overtone amplitudes.} Numerically evaluated overtone amplitudes $A_n$ in the small- and large-displacement case. The terms $A_0 = 1\cdot 10^{-7}$ and $A_1 = 3\cdot 10^{-7}$ are taken constant, with $F_\mathrm{o}$ corresponding to \SI{100}{\pico\newton}. For large displacement, the overtones have much higher amplitude than in the small-displacement case. The amplitude also decays more slowly with mode number.}
\label{Overtoneamplitudes}
\end{figure}

It is difficult to analytically express all the combinations of $j$ integers that sum up to $n$, so we numerically evaluate the overtone amplitudes in Fig.~\ref{Overtoneamplitudes}. Here, we take the expansion up to $13^\mathrm{th}$ order, which results in close to $2.5\cdot 10^6$ terms contributing to $n = 12$. We fix the terms with $n = 0,1$ to a constant value, in the experiment we do not directly apply a resonant drive ($A_1$) and there are radiation pressure effects that could result in a static position offset ($A_0$). From the numerical evaluations, we consistently see that for small displacement, the power in each subsequent overtone drops exponentially. However, for large displacement this is no longer true, and the higher-order terms from the expansion cause strong overtones that decrease much slower in amplitude with overtone number. This is reproduced in the experiments shown in the main text.

\subsubsection{Numerical simulation of overtones}
So far, we have derived analytically that the addition of an optical $\sin(x)$ non-linearity to a mechanical harmonic oscillator creates components of mechanical motion at integer multiples of the original frequency $\omega_0$. If the displacement is large enough, these overtones have significant amplitude and form a frequency comb. We have made some simplifying assumptions (e.g. $x_\mathrm{off} = 0$), which do not necessarily hold in practice. The remedy that, we perform numerical simulations. These allow us to better understand the roles of resonator displacement $x$, the the optical (dielectrophoretic) force $F_\mathrm{o}$, and the position offset $x_\mathrm{off}$ of the optical potential with respect to the mechanical potential.




\begin{figure}
\includegraphics[width = 0.5\textwidth]{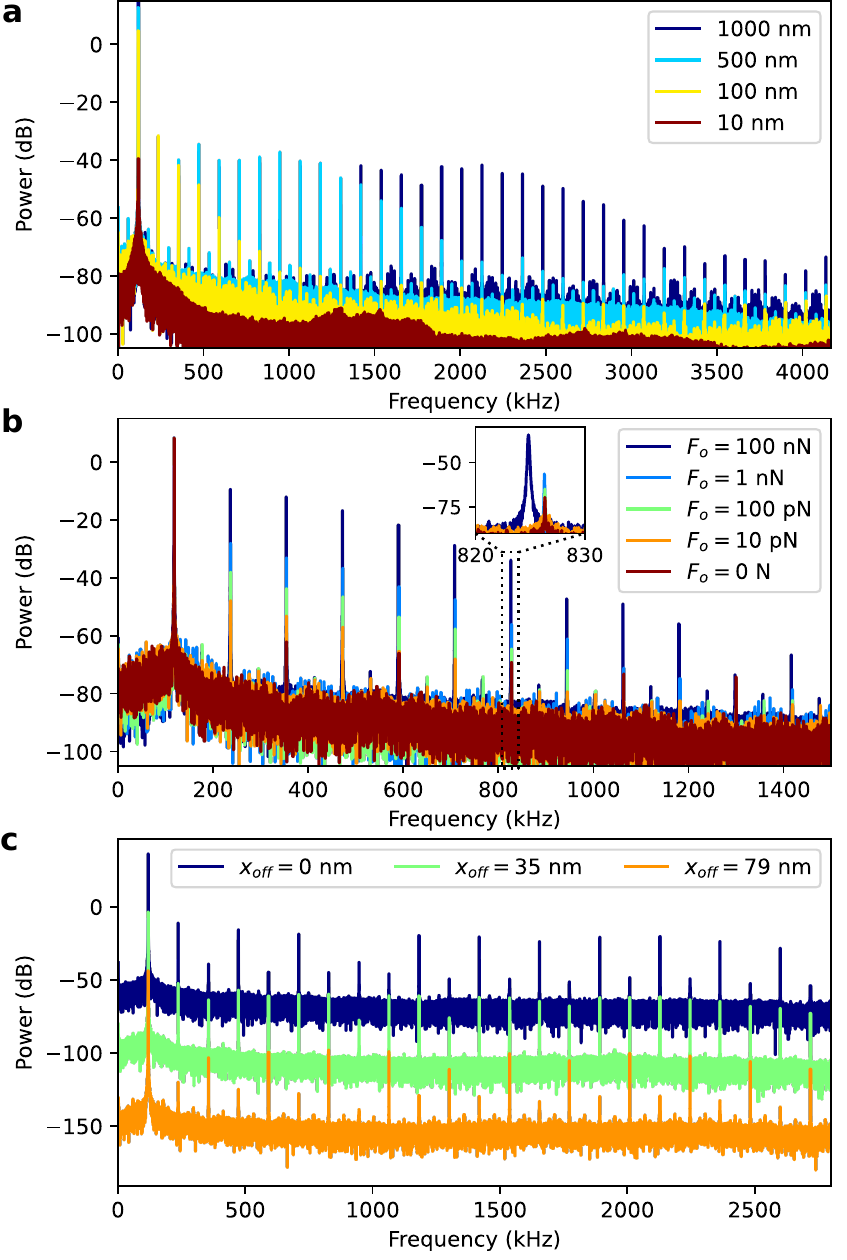}
\caption{\textbf{a}: Numerical simulation of overtones generated depending on initial amplitude. Parameters $F_\mathrm{o} = 100$~\si{\pico\newton}, $\omega_0 = 2\pi \times 118.049$~\si{\kilo\hertz}, $\gamma = 2\pi \times 0.02$~\si{\hertz}, $F_\mathrm{d} = 0$~\si{\newton}, $x_\mathrm{off} = 30$~\si{\nano\meter}. \textbf{b}: Simulation of overtones generated depending on optical force $F_\mathrm{o}$, for $x_o = 200$~\si{\nano\meter}. Other parameters identical to \textbf{a}. Inset shows the frequency shift for stronger optical fields. \textbf{c}: Simulation of overtones depending on offset $x_\mathrm{off}$. For $x_\mathrm{off} = 0$~\si{\nano\meter}, we predominantly drive the even overtones through the cosine-expansion, while for $x_\mathrm{off} = \lambda/8 = 79$~\si{\nano\meter} we predominantly drive the odd overtones through the sine-expansion. At $x_\mathrm{off} = 35$~\si{\nano\meter}, odd and even modes are approximately equal in power. Individual traces are offset vertically, $x_0 = 1000$~\si{\nano\meter} and other parameters are identical to \textbf{a}.}
\label{fignumericalsim}
\end{figure}

First, we investigate the dependence of the frequency comb on displacement $x$. In Fig.~\ref{fignumericalsim}\textbf{a}, we simulate the motion of the resonator starting from the value initial amplitude $x_0$ ($v_0 = 0$) indicated in the legend. We include dissipation but it is sufficiently small that the amplitude does not significantly decrease within the simulation time. For small amplitude of motion ($x_0 < x_\mathrm{off}$), we cross no optical extrema and only the fundamental mode is visible, no higher overtones appear. Once the motion is large enough to cross one optical extremum (\SI{100}{\nano\meter}), several overtones appear with exponentially decreasing power, as derived previously. If the motion is large enough to cross multiple optical extrema ($x_0 > \lambda/4 \simeq 150$~\si{\nano\meter}), more overtones appear. The lower overtones have approximately equal amplitude (empirically until $\omega = 3 x_0 / (\lambda/4) \times \omega_0$), since these also interact with the optical potential and get modulated to drive higher overtones. Note that the lower overtones have the same amplitude regardless of the resonator displacement. The higher overtones, above $\omega = 3 x_0 / (\lambda/4) \times \omega_0$, have an amplitude that decays exponentially. 

Secondly, we study the dependence of the frequency comb on dielectrophoretic force $F_\mathrm{o}$. For identical initial position $x_0$, we plot the frequency comb for various values of $F_\mathrm{o}$ in Fig.~\ref{fignumericalsim}\textbf{b}. For $F_\mathrm{o} = 0$~\si{\newton}, we see the dominant fundamental mode but also some higher harmonics (3,5,7,...). These are from the numerical accuracy of our simulation, they do not follow the exponentially decaying trend of the other curves. For increasing values of $F_\mathrm{o}$, the amplitude of each of the individual overtones increases linearly, but the exponential fall-off for higher overtone numbers is unchanged. At some point, the non-linearities start to shift the frequency of the modes, but this trap strength is well beyond the regime of our setup ($F_\mathrm{o} = 100$~\si{\nano\newton} corresponds to approximately \SI{6}{\watt} of incident laser power). 

Finally, we report on the dependence of the frequency comb on position offset $x_\mathrm{off}$. In the analytical case, we simplified using $x_\mathrm{off} = 0$~\si{\nano\meter}, retaining only the cosine terms so the even overtones are strong (dark blue). By shifting the optical potential with respect to the mechanical zero by $\lambda/8 = 79$~\si{\nano\meter}, we see mainly the odd-numbered overtones appear. For this value of $x_\mathrm{off}$, we can rewrite the cosine into a sine, such that the Taylor-series expansion only contains odd terms. At other values of $x_\mathrm{off}$, we have a weighted average of the two series expansions. In the middle, at $x_\mathrm{off} = 35$~\si{\nano\meter}, we see odd and even numbered modes with approximately equal power. 

To summarize: The addition of an optical $\sin(x)$ non-linearity to a mechanical harmonic oscillator creates components of mechanical motion at integer multiples of the original frequency $\omega_0$, which we call the overtones of $\omega_0$. These overtones can have significant amplitude, if the mechanical displacement is large enough. The displacement $x$ controls the number of overtones visible, while the dielectrophoretic force $F_\mathrm{o}$ determines their power relative to the fundamental mode of the comb. Based on the offset between the optical and mechanical potential $x_\mathrm{off}$, we can control the relative power in the even- or odd-numbered overtones. We have thus derived an analytical model for the overtone frequency comb, and expanded our qualitative understanding with numerical simulations.

\subsubsection{Exclusion of mechanical nonlinearity}
\begin{figure*}
\includegraphics[width = \textwidth]{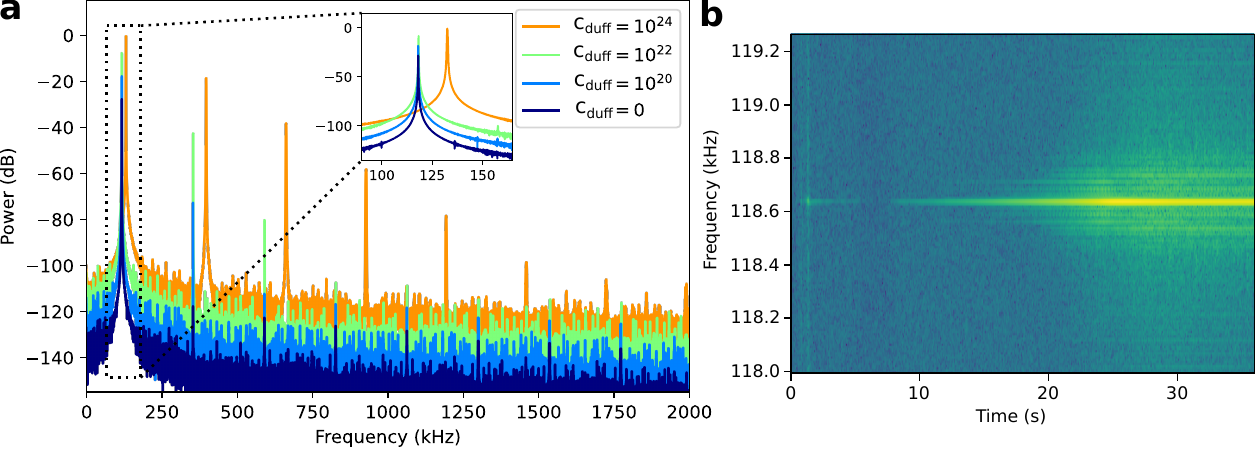}
\caption{\textbf{Frequency comb by mechanical nonlinearity. a}: Simulated displacement spectrum of a mechanical resonator with a Duffing nonlinearity. All traces share the same initial displacement, $x_0 = 500$~\si{\nano\meter}, with different Duffing coefficients. For a sufficiently large nonlinearity, the frequency comb appears, but the frequency of the fundamental mode (and comb spacing) is changed. \textbf{b}: Measured frequency of the fundamental mode of Fig.~3\textbf{e}, showing no shift in frequency between low displacement (no comb) and high displacement (comb).}
\label{ExcludeDuffing}
\end{figure*}

We have introduced an optical nonlinearity to explain the mechanical frequency combs, but we have not excluded conventional mechanical nonlinearities (e.g.~a Duffing term, $\propto x^3$). These nonlinearities can also lead to mechanical frequency combs, even in the absence of other modes to couple to~\cite{Yang2019,Ochs2022}. In Fig.~\ref{ExcludeDuffing}\textbf{a}, we have simulated and plotted the spectrum of a mechanical resonator with a Duffing term $c_\mathrm{duff} x^3$, such that the equation of motion is
\begin{equation}
\ddot{x} + \gamma \dot{x} + \omega_0^2 x + c_\mathrm{duff} x^3 = 0.
\end{equation}
We start from an initial condition of $x_0 = 500$~\si{\nano\meter} for all traces. For a sufficiently strong nonlinearity, we see a frequency comb. However, due to the nonlinearity, the frequency is shifted away from $\omega_0$. This is expected behavior for nonlinear resonators, oscillating at $\omega_0$ for small amplitude frequency shifting for larger amplitudes where the nonlinearity becomes dominant. The sign of $c_\mathrm{duff}$ controls the direction of the frequency shift, and a hardening nonlinearity (increasing frequency) is expected for most mechanical resonators.

In Fig.~\ref{ExcludeDuffing}\textbf{b}, we have plotted the fundamental mode during the measurement shown also in Fig.~3\textbf{e}. At the start of the measurement, the amplitude is small and there is no frequency comb, whereas at the end the amplitude is large and the comb is clear. There is no shift of the frequency of the fundamental mode visible in this measurement, which excludes the explanation of a Duffing nonlinear term being the origin of the comb.

It was proposed by others that strong, higher order nonlinearities (e.g.~$\propto x^3$, $x^5$, $x^7$, etc.) could be the origin of this comb~\cite{Yang2019}. However, to reproduce the frequency comb without a similar frequency shift as in the Duffing case, one would need strong nonlinear terms of very high orders. Rather than assuming as many nonlinear terms as we have overtones (up to $35$), we have introduced an optical nonlinearity that can explain all overtones from a single effect.

\subsubsection{Comparison to other combs}
There have been demonstrations of mechanical frequency combs generated via different mechanisms in literature. For applications, the comb properties are more important than the generation mechanism, and we have summarized a large part of the combs  available from literature in Table~\ref{Comparisoncombstable}. The entries are sorted by material/geometry platform, and multiple publications with subtle difference in the comb generation mechanism may be present within the individual categories.

The overtone comb of this work has a bandwidth on par with the largest available in literature (limited by the setup), but also has a relatively large resolution. The bandwidth is given by a combination of the maximum displacement and mechanical frequency, both can easily be engineered. We estimate the optothermal parametric drive is limited to $\lesssim 500$~\si{\kilo\hertz} in our membrane, which would constrain the mechanical frequencies that can be operated without an external drive. The frequency resolution could be enhanced by lowering the mechanical frequency (i.e.~make the membrane tethers longer and thinner), and can likely be tuned thermally. A further benefit is that the overtone combs are spectrally flat (same amplitude in different overtones), in contrast to other mechanical combs.

The main benefit of the overtone comb is that it does not require an external drive, which is unique for a mechanical comb. This greatly simplifies the necessary setup to operate the comb. In terms of uniformity of the frequency spacing, overtone combs come close to the level of electromechanical combs, where the uniformity is given by the electric drives directly. However, our measured uniformity is limited by the detection setup. Overtone combs are also among the most stable, especially comparing our 6-hour frequency stability (Sec.~\ref{Combstability}) to the \SI{10}{\second} Allan deviation of Ref.~\cite{Yang2021e}.

\begin{table*}
\begin{tabular}{l|ll|ll|l|l|l|l}
Type & \multicolumn{2}{l}{Bandwidth (\si{\hertz})} & \multicolumn{2}{l}{Resolution (\si{\hertz})} & Ext. drive power (dBm) & Uniformity (-) & Stability (-) & Footprint (\si{\micro\meter\squared}) \\
\hline
Electromechanical~\cite{Mahboob2012} & $20-100$ & & $0.005-10$ & & $-15$~\cite{Mahboob2016} & $1\cdot 10^{-9}$ & & $3\cdot 1 - 150\cdot 50$ \\
Nanostrings~\cite{Seitner2017} & $10-25$ & $(\times 10^3)$ & $0.5-10 $ & $(\times 10^3)$ &  $-2.5$~\cite{Ochs2022} & & & $55\cdot 0.27$ \\ 
Free-beam~\cite{Ganesan2017} & $60-150 $ & $(\times 10^{3})$ & $2-10 $ & $(\times 10^3)$ \cite{Ganesan2018} & $5$ & $5\cdot 10^{-6}$~\cite{Yang2021e} & $1\cdot 10^{-8}$~\cite{Yang2021e} & $1100\cdot 350$ \\
Coupled beams~\cite{Czaplewski2018} & $10-200$ & & $1-30$~\cite{Wang2022} & & $-26$ & & & $500\cdot 50$ \\
Bulk acoustic~\cite{Goryachev2020} & $20$ & & $0.7-2$ & & $-67$ & & & $23000 \cdot 1000$ \\
Circular 2D~\cite{Chiout2021,Keskekler2022} & $2-11 $ & $(\times 10^6)$ & $75-400 $ & $(\times 10^3)$ & $10$ & & & $5-8$ (diameter) \\
This work & $4.2 $ & $(\times 10^6)$ & $118-359 $ & $(\times 10^3)$ & None & $4.7\cdot 10^{-8}$ & $7.5\cdot 10^{-10}$ & $750\cdot 750$
\end{tabular}
\caption{\textbf{Mechanical frequency comb performance.} Overview of reported mechanical frequency comb properties from literature, organized by platform (irrespective of comb mechanism). The overtone comb has a bandwidth (frequency span) on par with the largest combs, but also a relatively large resolution (frequency spacing). We report the absolute frequency stability over a 6-hour period (see SI Sec.~\ref{Combstability}), whereas \cite{Yang2021e} computes the Allan deviation over a \SI{10}{\second} period. Our work is the only mechanical frequency comb that does not require any external drive.}
\label{Comparisoncombstable}
\end{table*}

\subsection{Higher harmonics in LDVs}\label{Higherharmonics}
There are several mechanisms by which spurious higher harmonics can appear in measurements from a LDV~\cite{Siegmund2008}. In the following paragraphs, we will discuss these mechanisms and show that we can exclude them as the source of the observed frequency comb. We will treat all parts of the setup shown in Fig.~5: the optical parts, the photodetector, the decoder and the data acquisition.

\begin{figure}
\includegraphics[width = 0.5\textwidth]{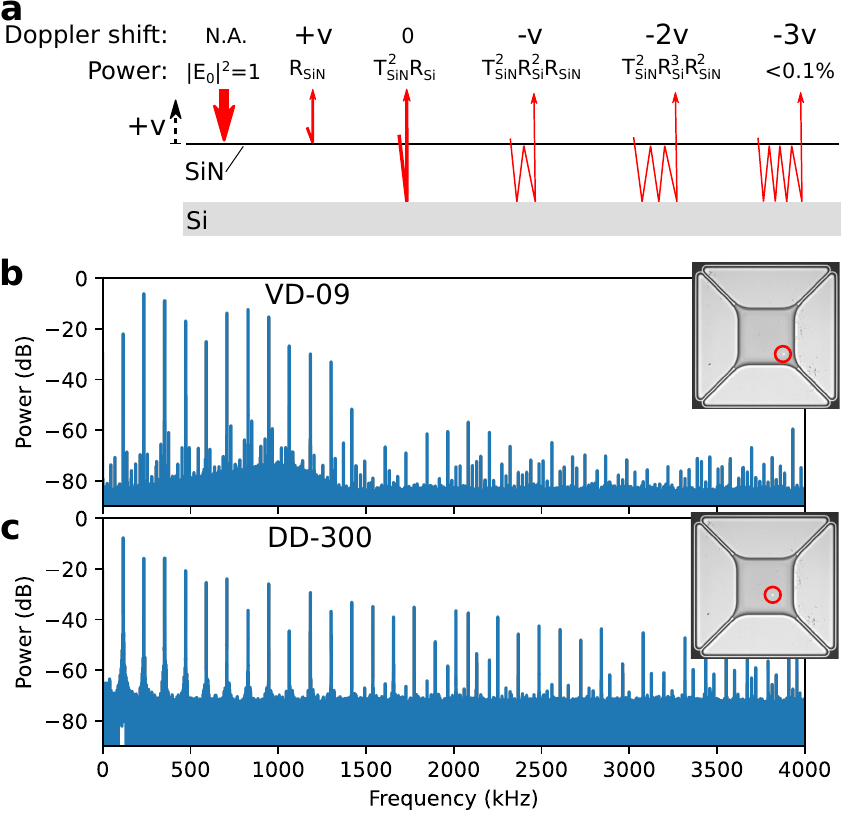}
\caption{\textbf{Sources of higher harmonics. a}: Schematic of reflected components of a single optical beam on our device (red arrows). For each component, the Doppler shift due to membrane velocity $v$ is shown, as well as the relative power. \textbf{b}: Frequency comb measured using the VD-09 decoder (maximum bandwidth set to \SI{1.5}{\mega\hertz} to allow maximum detectable amplitude). Inset shows microscope image with laser position highlighted by the red circle. \textbf{c}: Frequency comb measured using the DD-300 decoder, measured on the same device at a slightly different position.}
\label{Comparisoncombs}
\end{figure}

Multi-path interferences may happen when the LDV laser beam is reflected by more than one surface~\cite{Yarovoi2004}. The geometry of our system facilitates having multiple reflection sources, since both the trampoline membrane and the substrate contribute to the reflection. We assume an ideal LDV, with an incident electrical field with amplitude $E_0$, a moving membrane, and a stationary substrate, as shown in Fig.~\ref{Comparisoncombs}\textbf{a}.  The directly reflected (desired) beam from the membrane has power  $R_\mathrm{SiN} |E_0|^2$ ($R_\mathrm{SiN} = 0.3$). Assuming no scattering ($T + R = 1$, for transmission $T$ and reflection $R$), the substrate contributes $T_\mathrm{SiN}^2 R_\mathrm{Si} |E_0|^2 \simeq 0.17 |E_0|^2$ ($R_\mathrm{Si} = 0.35$). Only the membrane contribution has a Doppler shift due to its motion, $+v$. Neglecting effects from beam divergence, multiply-reflected beams contribute $T_\mathrm{SiN}^2 R_\mathrm{Si}^2 R_\mathrm{SiN}$, $T_\mathrm{SiN}^2 R_\mathrm{Si}^3 R_\mathrm{SiN}^2$, and so on. These terms have Doppler shifts with the opposite sign to the directly reflected beam, as they interact with the membrane from the opposite direction. The first term has a Doppler shift equal in magnitude to the desired signal ($-v$), while subsequent terms have multiple times the Doppler shift ($-2v, -3v, ...$). These terms thus directly affect the magnitude of the velocity that the LDV senses. For harmonic membrane motion $v(t) = v_0 \cos(\omega_0 t)$, the multiply-reflected beams lead to an error in $v_0$, but their Doppler shift does not contribute directly to higher harmonics of the observed membrane motion.

In a non-ideal laser Doppler vibrometer, multi-path interferences can be identified by presence of ripples and spikes in the velocity signal. We can compare the measured velocity signals with simulated signals using the model of~\cite{Yarovoi2004}. For a harmonic oscillator with $\omega_0 = 118$~\si{\kilo\hertz}, maximum displacement amplitude $x_\mathrm{max}$, the demodulated velocity signal has the form
\begin{equation}
v(t) = -\frac{1}{2}\omega_0 x_\mathrm{max} \left( \frac{\theta_2 - 1}{2\theta \cos\left(\frac{4\pi}{\lambda} \cos(\omega t) + \Delta\phi \right) + \theta^2 + 1}\right)\sin(\omega t).
\end{equation}
Here, $\theta$ is the electric field ratio of the correct and the unwanted reflected signals ($\theta = \infty$ for an ideal vibrometer), $\Delta\phi$ is the phase offset between the two beams (constant). In the worst-case scenario ($R_\mathrm{SiN} = 0.3$, $R_\mathrm{Si} = 0.35$, no scattering losses), $\theta \simeq 1.33$. The unwanted beam likely suffers from scattering losses more than the correct beam, since its path crosses the Si$_3$N$_4$ domain twice and reflects of the etch-roughened Si surface (see Fig.~5\textbf{d}). 

\begin{figure*}
\includegraphics[width = 0.8\textwidth]{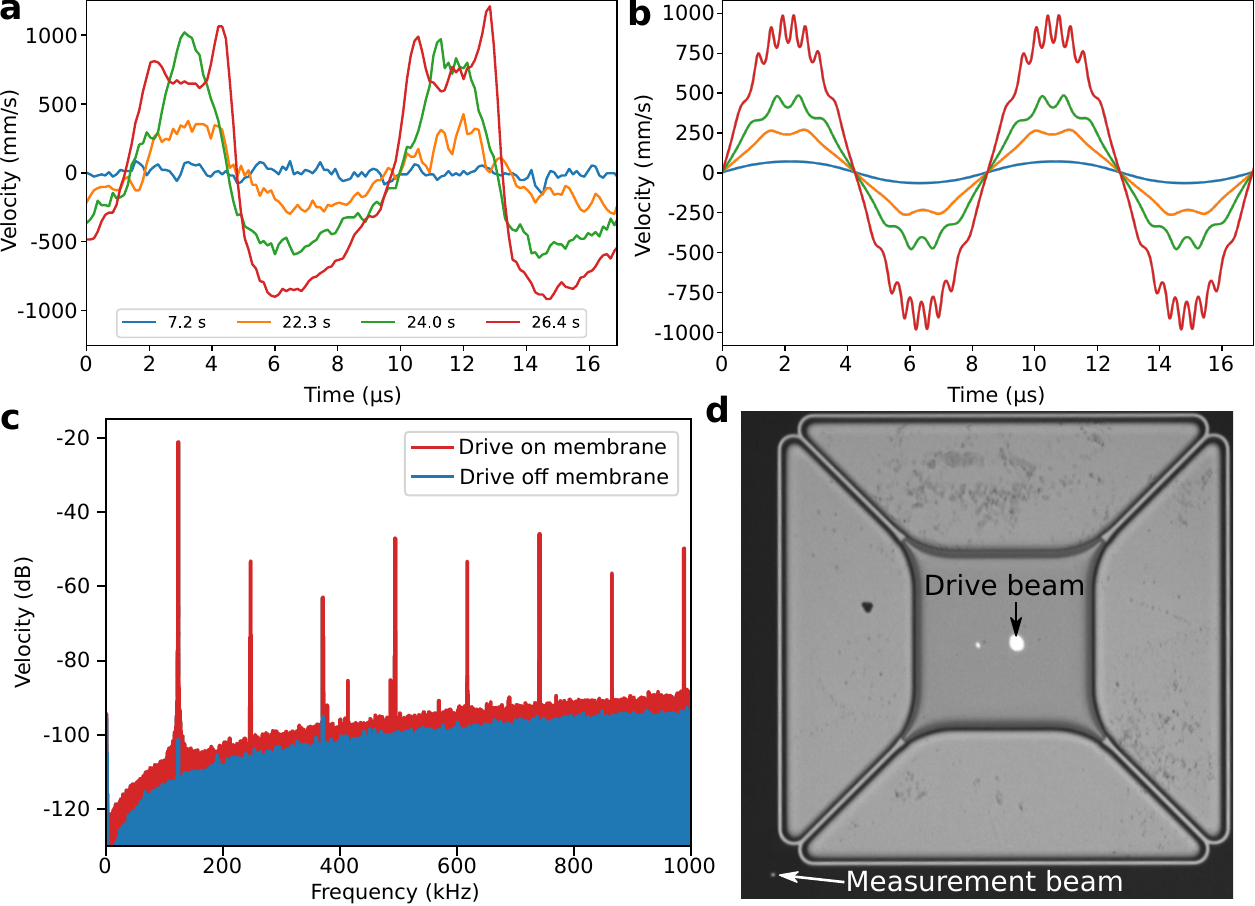}
\caption{\textbf{Exluding multi-path interferences. a}: Measured velocity signal for different times in Fig.~4\textbf{b}. The data has been smoothed with a first order Savitzky-Golay filter of width $3$, to emphasize the presence of ripples. \textbf{b}: Simulated velocity signal of a resonator with frequency \SI{118}{\kilo\hertz} with multi-path interferences present. The characteristic ripples of the interference are visible at the extrema of the velocity curve in \textbf{b}, but absent in \textbf{a}. \textbf{c}: Measured velocity spectrum outside the membrane, on chip. A second (drive) beam generates a frequency comb of the membrane's fundamental mode, which we detect at the position indicated on the microscope image \textbf{d}. Without this beam on the membrane, no comb is observed.}
\label{MultipathExclude}
\end{figure*}

We simulate the velocity signal with multi-path interferences, and compare it with the observed velocity signal in Fig.~\ref{MultipathExclude}. The simulated signal of Fig.~\ref{MultipathExclude}\textbf{b} ($\theta = 10$) has the characteristic ripples of multi-path interference. By comparing the simulated signal to the measured signal (Fig.~\ref{MultipathExclude}\textbf{a}), one sees that the characteristic ripples are absent. While it is unlikely that the multi-path interference effect is completely absent in our setup, it can be corrected by a combination of amplitude- and phase-locked loops~\cite{Dussarrat1998}. Either way, the observed velocity signal does not correspond to multi-path interference effects.

We can further exclude multi-path interferences as the source of the comb by measuring outside the membrane, on the Si chip. At the position indicated on the microscope image of Fig.~\ref{MultipathExclude}\textbf{d}, the Si and Si$_3$N$_4$ layers are touching and thus have the same motion. The multi-path interference as described in Ref.~\cite{Yarovoi2004} does not happen. We use a second beam ('drive beam' in Fig.~\ref{MultipathExclude}\textbf{d}) to generate a comb of the fundamental mode of the membrane. At the measurement position outside the membrane, we can detect this comb, red curve in Fig.~\ref{MultipathExclude}\textbf{c}. For verification, without the drive beam present (blue curve), no comb is detected while the fundamental mode is visible. Thus we can exclude multi-path interference effects as the source of the observed frequency combs.

High-aperture effects can occur when the LDV is operated with a microscope objective, based on the Guoy phase delay of the optical beam~\cite{Siegmund2008,Sumali2008}. The majority of the measurements in this work were done using a Mitutoyo Plan APO 5x objective with a numerical aperture of $0.14$. With this numerical aperture, the amplitude error and harmonic distortion should be limited to $<1\%$~\cite{Siegmund2008}. Additionally, we compare the frequency combs obtained from the same device using different lenses (different numerical apertures) in Sec.~\ref{Comblaser}, which show no relation between the numerical aperture and frequency comb. Thus we can exclude high-aperture effects as the source of the observed frequency comb.

The electronic parts of the LDV (sketched in Fig.~5\textbf{a}) may contribute to a nonlinear response that would generate harmonics~\cite{Siegmund2008}. We record frequencies well below the \SI{20}{\mega\hertz} specified maximum frequency of the detector, thus this component should not contribute higher harmonics.

In a conventional heterodyne interferometer, harmonics of a signal would naturally arise for sufficiently large displacement. The interferometer phase is linear only in a limited regime, typically $x \ll \lambda$. However, a Doppler vibrometer measures the frequency shift of the reflected light rather than its phase. The frequency shift can be extracted with a frequency-to-voltage converter, in the LDV decoder. This allows an LDV to measure velocities associated with displacements much larger than the optical wavelength, without suffering from the harmonics expected in a displacement interferometer.

There are two LDV decoders available in our setup, the VD-09 velocity decoder and the DD-300 displacement decoder. The VD-09 has a lower maximum operating frequency ($1.5-2.5$~\si{\mega\hertz}), while the DD-300 has a lower detection maximum. In Fig.~\ref{Comparisoncombs}\textbf{b,c}, we show the frequency combs measured from the same device with the different decoders. These measurements show that the frequency comb is not a decoder artifact. Thus we can exclude the electronic parts of the measurement setup as a source of the observed frequency comb.

In summary, we have described common mechanisms that yield higher harmonics in laser Doppler vibrometer measurements. All these mechanisms can be excluded as the cause of the frequency combs, by qualitative arguments and the measurements shown in Figs.~\ref{Comparisoncombs}, \ref{MultipathExclude}, and ~\ref{Objectivechange}. In combination with the fact that there is a visible change in the membrane when the frequency comb occurs (Fig.~\ref{Structuralchange}), we conclude that the observed mechanical frequency comb is a real, physical phenomenon and not an artifact of the measurement setup.

\subsection{Membrane in an optical trap}\label{Opticaltrap}
\begin{figure*}
\includegraphics[width = \textwidth]{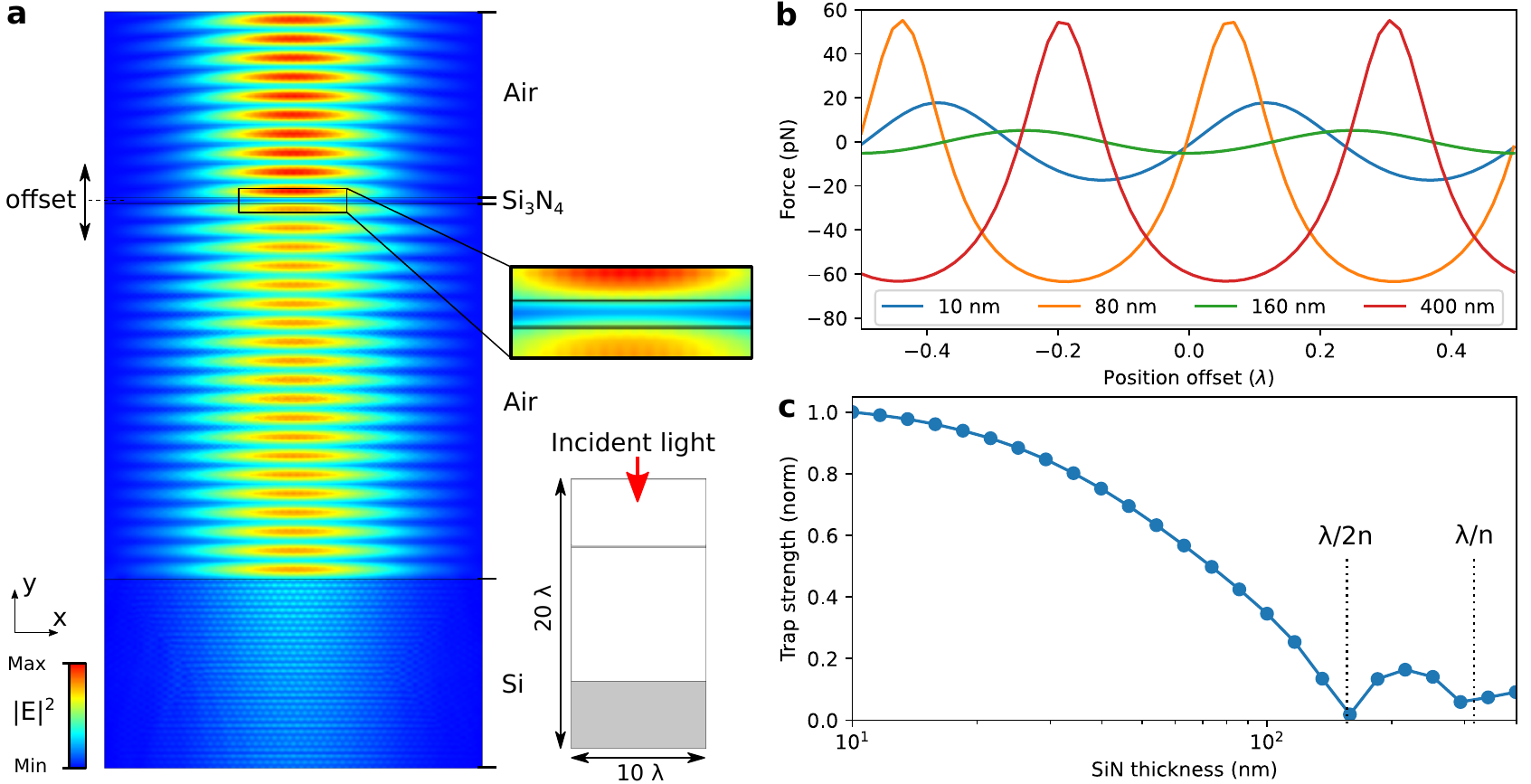}
\caption{\textbf{Simulation of optical trap. a:} Electric field norm of incident \SI{633}{\nano\meter} light on \SI{100}{\nano\meter} Si$_3$N$_4$ membrane suspended $\sim$\SI{6}{\micro\meter} above a Si backplane. The reflected light forms a standing wave pattern. Zoom shows center of the Si$_3$N$_4$ membrane. Bottom inset shows the simulation geometry schematically. \textbf{b:} Force exerted by the optical field on the Si$_3$N$_4$ domain as the position of the membrane is varied. The colors represent different membrane thicknesses. \textbf{c:} Normalized strength of the trap for different Si$_3$N$_4$ thicknesses.}
\label{figOpticaltrap}
\end{figure*}
Optical traps can be used to confine small particles, as the light exerts a force proportional to the gradient of the optical intensity~\cite{Ashkin1970,Ashkin1986}. For uncharged dielectric particles, the electrical field polarizes the particles, and any change in field (from a gradient) will result in a dielectrophoretic force acting on that particle~\cite{Pohl1951}. In this section, we will simulate a laser incident on a dielectric (Si$_3$N$_4$) membrane, and evaluate the dielectrophoretic and radiation pressure forces. We will show that the optical gradient is sufficient for the dielectrophoretic force to affect the mechanical motion, while the radiation pressure force does not affect the dynamics significantly.

\subsubsection{Dielectrophoretic force}
We simulate the physical system consisting of a $t\simeq$~\SI{80}{\nano\meter} thin Si$_3$N$_4$ slab suspended in air above a much thicker Si slab, shown in Fig.~\ref{figOpticaltrap}\textbf{a}. We study a 2D-axisymmetric domain centered around our laser spot, which is incident from the top of the figure. We assume that our laser Doppler vibrometer emits a Gaussian beam for convenience. For a regular optical trap, the optical gradient and thus the force exerted is weakest along the laser propagation direction, while it is stronger in the plane normal to the propagation direction. However, the Si backplane in our geometry reflects part of the light and produces a standing wave (Fig.~\ref{figOpticaltrap}\textbf{a}), producing a periodic optical intensity. Combined with the fact that the membranes are much larger than the optical gradient laterally ($\sim 750\times750$~\si{\micro\meter\squared}), we have a one-dimensional counterpropagating-wave optical trap in the out-of-plane direction. A simple (unpatterned) Si backplane has been shown before to be sufficient to create an interference pattern for such an optical trap~\cite{Zemanek1999}.

We consider only motion in the out-of-plane direction of the membrane, so we are interested in the gradient of the electric field norm in the $y$-direction (Fig.~\ref{figOpticaltrap}\textbf{a}). The total force $F_\mathrm{de}$ induced by the optics on the dielectric Si$_3$N$_4$ structure can then be found by integrating this gradient over the membrane domain,
\begin{equation}
F_\mathrm{de} = \int_{V_{SiN}} \alpha \frac{\partial |E|^2}{\partial y} \mathrm{d}V,
\label{Fototal}
\end{equation}
with $\alpha$ the polarizability. For a linear isotropic material, the constant $\alpha = P_x/E_x$ can be found from the simulation by dividing the polarization in any direction $P_x$ by the electric field in that direction $E_x$. 

We need relate the equation for $F_\mathrm{de}$ (units of force) to the coefficient $F_\mathrm{o}$ (units of force) used in Eq.~1. The expression for $F_\mathrm{de}$ depends on the position of the resonator with respect to the Si backplane, Fig.~\ref{figOpticaltrap}\textbf{b}, whereas $F_\mathrm{o}$ is a constant. $F_\mathrm{de}(x)$ in Fig.~\ref{figOpticaltrap}\textbf{b} shows the periodic behavior expected from an optical trap (i.e. periodic sign change), as well as a slight asymmetry due to radiation pressure. It shares the periodicity that we have separately introduced in Eq.~1 in the main text, $\sin\left(\frac{4\pi}{\lambda}(x-x_\mathrm{off})\right)$, but does not exactly follow the same curve due to the finite size (thickness) of the membrane with respect to the wavelength. To simplify the dynamical calculations, we have thus chosen to separate out the periodic behavior and leave it proportional to the optical intensity, $\sin\left(\frac{4\pi}{\lambda}(x-x_\mathrm{off})\right)$, and keep $F_\mathrm{o}$ as a constant equal to the maximum of $|F_\mathrm{de}(x)|$. Thus for a $t = $~\SI{80}{\nano\meter} membrane subject to our \SI{3.6}{\milli\watt} laser, we expect a maximum out-of-plane force in the order of \SI{60}{\pico\newton}. This confirms the values of $F_\mathrm{o}$ used in the numerical simulations in the main text and in Sec.~\ref{Overtones}, which show that the dynamics of our membrane can be significantly altered by the presence of the optical field.

The different curves of Fig.~\ref{figOpticaltrap}\textbf{b} illustrate that the total optical force exerted on the membrane does not decrease monotonically with membrane thickness. When we calculate the effective strength of the trap (peak-to-peak force difference divided by the thickness of the membrane, in Fig.~\ref{figOpticaltrap}\textbf{c}), we see that it decreases for increasing Si$_3$N$_4$ thickness. There are minima around $t = \lambda/(2n_\mathrm{SiN})$ where $n_\mathrm{SiN}$ the index of refraction; the half-integer number of optical periods leads to a total cancellation of the net force on the membrane $F_\mathrm{de}$. 

\subsubsection{Radiation pressure force}
In an optical cavity, the interaction between radiation pressure and a mechanically compliant element can lead to optomechanical frequency combs~\cite{Miri2018}. The radiation pressure of an optical cavity coupled to a mechanical resonator can result in an effective cubic non-linearity, and by sufficiently driving the optical cavity one can bring the system in the self-oscillation regime. This creates an optical frequency comb with a spacing equal to the mechanical frequency. We must thus distinguish the frequency combs from the dielectrophoretic force from those originating from the radiation pressure.

We follow the analysis of Miri et al.~\cite{Miri2018} as they analyze the stability of a mechanically compliant end-mirror optomechanically coupled to an optical cavity. Our Si$_3$N$_4$ membrane functions as the compliant end mirror, while the Si substrate is the other (fixed) end mirror. For a given set of parameters, we evaluate whether the frequency comb will appear through equations (9a,b) of \cite{Miri2018}, which amounts to checking the Routh-Hurwitz stability criterion. For completion, the conditions for stability are
\begin{equation}
\begin{aligned}
3\mathcal{G}^2 + 4\Delta\mathcal{G} + \Delta^2 + \kappa^2/4 &>0, \\
\mathcal{G}^4 + d_1 \mathcal{G}^3 + d_2 \mathcal{G}^2 + d_3 \mathcal{G} + d_4 &> 0,
\end{aligned}
\end{equation}
with coefficients
\begin{equation}
\begin{aligned}
d_1 = &4\Delta, \\
d_2 = &\gamma\kappa + 6\Delta^2 + \gamma^2 -6\omega_0^2 + \frac{\kappa^2}{2} - \frac{2\gamma\omega_0^2}{\kappa} - \frac{2\omega_0^2\kappa}{\gamma}, \\
d_3 = &\Delta \left( 2\gamma \kappa + 4\Delta^2 +2 \gamma^2 -8\omega_0^2 + \kappa^2 - \frac{2 \gamma\omega_0^2}{\kappa} - \frac{2\omega_0^2 \kappa}{\gamma} \right), \\
d_4 = &\Delta^2 \left( \Delta^2 + \gamma^2 +\gamma \kappa - 2 \omega_0^2 + \frac{\kappa^2}{2}  \right) + \frac{\gamma^2 \kappa^2}{4} \\ 
& + \gamma\omega_0^2 \kappa + \frac{\gamma\kappa^3}{4} + \left( \omega_0^2 + \frac{\kappa^2}{4} \right)^2.
\end{aligned}
\end{equation}
The key quantity is the static optomechanical frequency shift $\mathcal{G} = \frac{2 g_0^2}{\omega_0} |a|^2$, with $g_0 = 2\pi \times 1$~\si{\hertz} the vacuum optomechanical coupling strength typical for these membranes~\cite{deJong2022a}. Additionally, $a = \left(\frac{\kappa P_\ell/4\hbar\omega_\ell}{(\kappa/2 + i\Delta)}\right)^{1/2}$ is the average optical field amplitude in the cavity, determined by the cavity decay rate $\kappa$, laser power and frequency $P_\ell, \omega_\ell$ and detuning $\Delta$. Finally, the mechanical frequency $\omega_0 = 2\pi \times 120$~\si{\kilo\hertz} and linewidth $\gamma = 2\pi \times 0.1$~\si{\hertz} are rounded from the values measured and reported in the main text.

To estimate $\kappa$, we calculate the finesse of the Fabry-Pérot cavity formed by a \SI{10}{\micro\meter} distance between an $R = 0.3$ Si$_3$N$_4$ and an $R = 0.35$ Si backplane. This is a low estimate for $\kappa$, as the Si surface is rough from fabrication (Fig.~5\textbf{d}), and we do not consider additional losses from that. We obtain $\kappa \sim 2\pi\times 430$~\si{\giga\hertz}. We assume the optimal detuning to reach the frequency comb regime, which is around $\Delta = - \kappa/4$. Combining these values, we can evaluate the expressions for the stability and determine a minimum input laser power to reach the frequency comb regime. This is at approximately \SI{1.5}{\kilo\watt}, several orders of magnitude larger than the \SI{3.6}{\milli\watt} measured laser power output of our laser Doppler vibrometer. Thus the optomechanical (radiation pressure) instability is not the origin of our frequency combs.

In conclusion, we have simulated the behavior of a Si$_3$N$_4$ membrane subject to an incident laser. Due to the Si backplane, reflected light forms a standing wave which leads to a counterpropagating optical trap. We have calculated the force exerted on the membrane by the optical trap, its dielectrophoretic part is sufficient to generate the overtone frequency comb as described in the main text. We have also verified that the optical cavity formed between the Si$_3$N$_4$ membrane and Si backplane is not good enough to lead to an optomechanical frequency comb based on the radiation pressure.

\subsection{Optothermal parametric driving}\label{Thermalparametricdriving}
\begin{figure*}
\includegraphics[width = \textwidth]{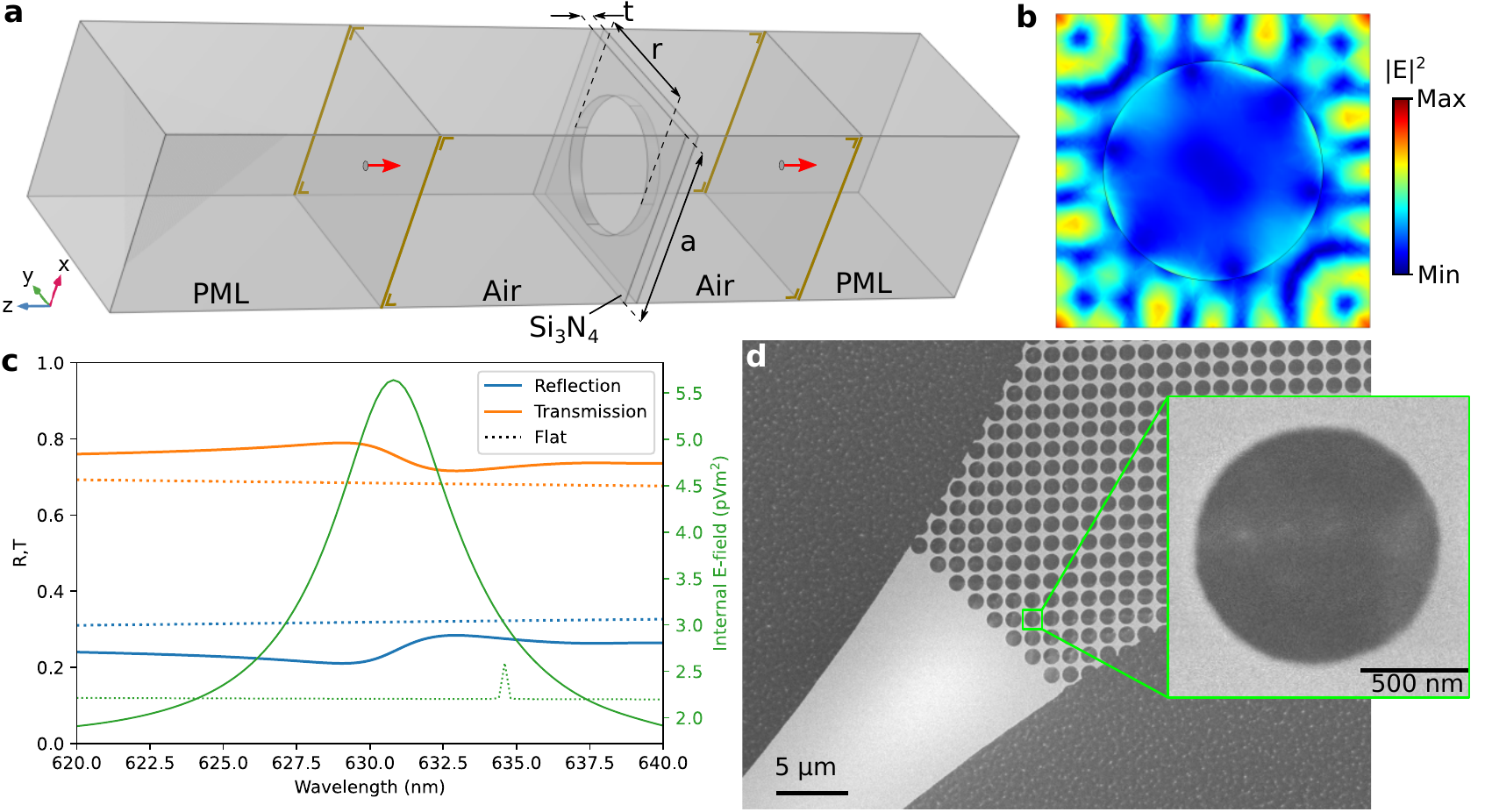}
\caption{\textbf{Optical behavior of photonic crystal. a:} Schematic of simulation setup, consisting of a Si$_3$N$_4$ photonic crystal unit cell in air, with a plane wave incident in the out-of-plane direction (red arrows). There are periodic boundary conditions on the outer surfaces, with the brown markings indicating one of the surface pairs. \textbf{b:} Norm of the E-field along the mid-surface of the Si$_3$N$_4$ domain, showing a standing wave for $\lambda = 630.8$~\si{\nano\meter}. \textbf{c:} Reflection and transmission of the photonic crystal as a function of wavelength, and that of a flat Si$_3$N$_4$ surface for comparison (dotted lines). The feature in the reflection/transmission is correlated with the mode of the photonic crystal shown in \textbf{b}, indicated by the total E-field integrated over the Si$_3$N$_4$ domain (green). This feature is also absent for an unpatterned surface (green, dotted). \textbf{d}: SEM image of photonic crystal at the membrane edge, with inset showing single hole.}
\label{PhCOptical}
\end{figure*}

Driving the trampoline resonators to sufficiently large amplitudes to see the frequency comb from the optical trapping potential requires a strong driving mechanism. We have previously used a piezo shaker to drive these membranes~\cite{deJong2022b} (also described in Sec.~\ref{Combwithpiezodriving}), but for the majority of this work we use a optothermal parametric drive mechanism where the membrane reaches the self-oscillation regime~\cite{Aubin2004}. That means that we can use a continuous-wave laser without power or frequency modulation, and do not require any additional signal or connection to the chip containing the resonator. 


In the following subsections, we study the optothermal parametric driving. Since it is based on absorption, which is normally not too large for $\lambda = 633$~\si{\nano\meter} in Si$_3$N$_4$, we first show the absorption is enhanced by our photonic crystal structure. Then, we simulate the slow ($>0.1$~\si{\second}) thermal dynamics from steady state absorption and the fast ($\ll 0.1$~\si{\second}) thermal dynamics due to modulated intensity from the standing wave optical field. Finally, we compare the simulated changes in temperature to the optothermal self-oscillation theory~\cite{Aubin2004} and other effects from literature.

\subsubsection{Absorption in the photonic crystal}
Si$_3$N$_4$ is a often-used optical material because of its low absorption at telecom wavelengths, with typical absorption coefficients $\mu = 1-2$~\si{\per\centi\meter} at \SI{1550}{\nano\meter}~\cite{Steinlechner2017}. However at \SI{633}{\nano\meter}, absorption is higher and sensitive to deposition parameters (reports vary from $\mu = 2$~\si{\per\centi\meter}~\cite{Frigg2019} to $\mu = 5000$~\si{\per\centi\meter}~\cite{Gorin2008}. Works using a similar setup but low-stress Si$_3$N$_4$ measured \SI{0.5}{\percent} absorption for a \SI{50}{\nano\meter} film ($\mu = 100$~\si{\per\centi\meter})~\cite{Chien2018}, so for simulation purposes we assume \SI{1}{\percent} for our \SI{80}{\nano\meter} film. It is difficult to verify the accuracy of this assumption, especially since our trampoline resonators are patterned with an array of circular holes that form a photonic crystal~\cite{Gaertner2018} and previously cited absorption coefficients cover unpatterned films. These works suggest absorption should be dependent mainly on material properties (Si$_3$N$_4$ deposition parameters) thus be the same for all membranes on a chip. However, we experience quite a variance in the response of our membranes to our laser beam (SI Sec.~\ref{Comblaser}), and also depending on the exact positioning of the laser on the membrane, so we propose a different effect.

Photonic crystal patterns can be designed to yield a certain reflectivity at a target wavelength~\cite{Gaertner2018}, but for the resonators in this work the photonic crystal functions as a mechanism to accurately control the resonance frequency~\cite{deJong2022b}, as well as allowing their smooth undercut. Due to this, the photonic crystal parameters (lattice spacing $a = 1356$~\si{\nano\meter}, hole radius $r = 481$~\si{\nano\meter}) are larger than the wavelength, which allows standing waves in the Si$_3$N$_4$ domain that couple to waves incident on the plane of the membrane. These modes enhance the field inside the Si$_3$N$_4$ and thus could increase the absorption while being subject to fabrication imperfections such that not all membranes or locations on membranes absorb equally, leading to the variance in the observed optothermal self-oscillation. 

We simulate the behavior of the photonic crystal in COMSOL by taking a unit cell with the photonic crystal parameters ($a$, $r$, and Si$_3$N$_4$ thickness $t = 100$~\si{\nano\meter}), as shown in Fig.~\ref{PhCOptical}\textbf{a,d}. We use $n=2.016$ for the Si$_3$N$_4$ domain and $n=1$ for vacuum. Parallel to the plane of the membrane we define two periodic ports, for input (reflection) and output (transmission), which are located sufficiently far away from the Si$_3$N$_4$ domain. These ports border a perfectly-matched layer that absorbs scattered light that is not absorbed by the periodic ports. The incident wave is an plane wave with normal incidence to the slab, with a polarization in the $x+y$ direction, and the entire simulation domain has periodic boundary conditions in $x$ and $y$ directions (brown lines in Fig.~\ref{PhCOptical}\textbf{a}. 

From this simulation, we find that a standing wave pattern forms in the plane of the Si$_3$N$_4$ membrane, shown in Fig.~\ref{PhCOptical}\textbf{b}. When we evaluate the reflection and transmission of this photonic crystal, we see only a small feature around $\lambda = 630.8$~\si{\nano\meter}, Fig.~\ref{PhCOptical}\textbf{c}. However, the field integrated over the Si$_3$N$_4$ domain is much enhanced. These modes are sensitive to material thickness, $a$ and $r$, and not unique to the parameters chosen for our simulation. We propose that standing waves of this form cause strong(er) absorption of light in some of our structures, depending on fabrication imperfections and precise laser position. 

\subsubsection{Slow thermal behavior}\label{Slowthermal}
\begin{figure}
\includegraphics[width = 0.5\textwidth]{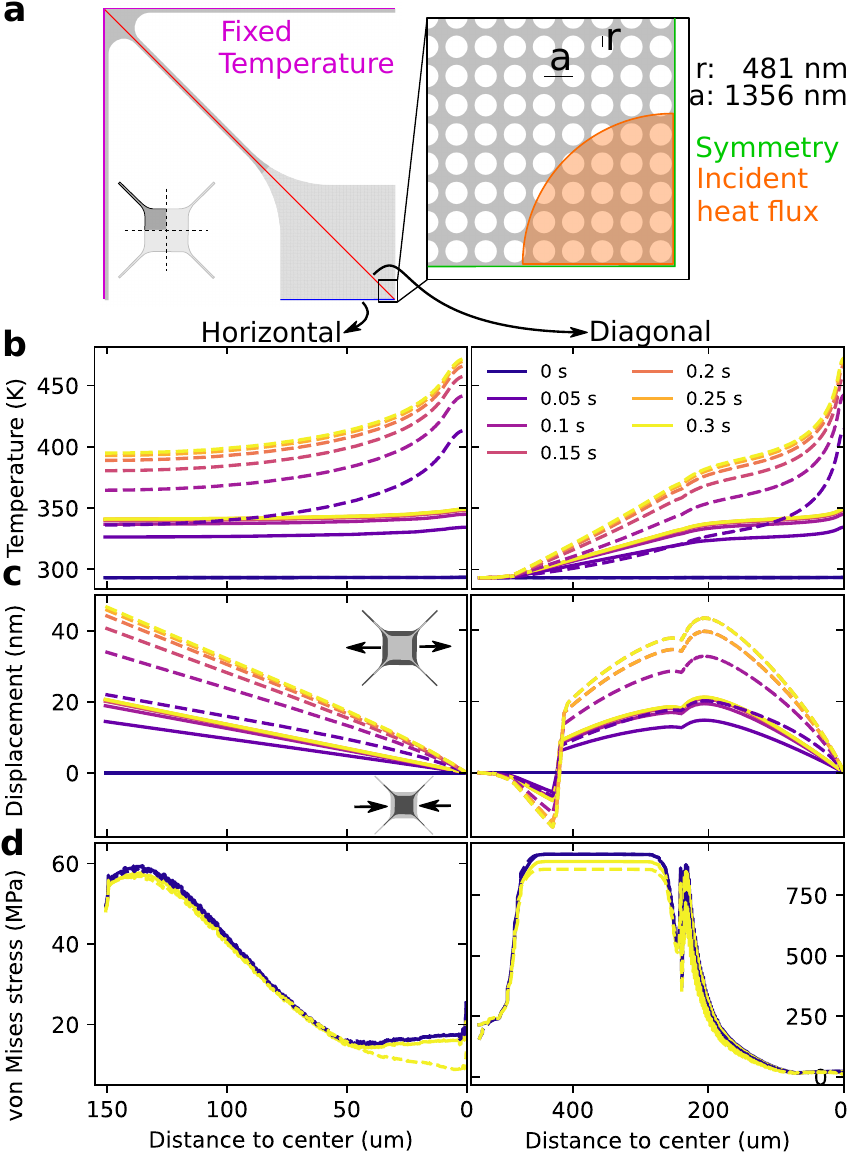}
\caption{\textbf{Thermal behavior of photonic crystal. a:} Schematic of simulation of the laser heating the suspended membrane. Boundary conditions and geometry are described in the text. Results are extracted along two linecuts, horizontal (left column) and diagonal (right column) in \textbf{b, c, d}. Starting from room temperature, an incident heat flux representing absorption from the laser light heats up the membrane over time. Colors indicate time, solid lines assume thermal conductivity $k =$~\SI{20}{\watt\per\meter\per\kelvin}, dashed lines use $k =$ ~\SI{2}{\watt\per\meter\per\kelvin}. \textbf{b:} Temperature along the linecuts shows heating in the center, which settles within \SI{0.3}{\second}. \textbf{c} Displacement of the geometry away from (positve) or towards (negative) the center of the domain due to thermal expansion. The maximum displacement is $\sim 45$~\si{\nano\meter}. \textbf{d:} Von Mises stress (smoothed, only first and last timestep shown). Left and right panel have different y-axis scale.} 
\label{PhCthermal}
\end{figure}
To build our understanding of the effects of the absorbed light, we simulate the effect of a heat source on the photonic crystal membrane (Fig.~\ref{PhCthermal}\textbf{a}). The membrane structure is symmetric, so we can reduce the simulation domain to only a quarter of a membrane. The outer edges of the domain connect the suspended Si$_3$N$_4$ connects to the (much more massive) Si substrate, so we apply a fixed temperature condition (purple). The heating due to the laser light is considered as an incident heat flux in the center of the membrane (orange), defined via the total deposited heat as $P\mu_\mathrm{Si_3N_4}$ with laser power $P = 3.6$~\si{\milli\watt} and absorption coefficient $\mu_\mathrm{Si_3N_4} = 0.01$. The heat can be lost through the fixed temperature boundary condition at the outer edges of the domain, but also through radiative transfer in the out-of-plane direction, where we assume the environment is at room temperature. We use a surface emissivity of $0.1$, which is valid for room temperature and a film thickness of \SI{100}{\nano\meter}~\cite{Zhang2020PRA}, and noticeably different than the $0.6-0.9$ value used for bulk Si$_3$N$_4$.


The choice of material parameters of Si$_3$N$_4$ is important for our simulation. The Young's Modulus $E = 250$~\si{\giga\pascal}, Poisson's ratio $\nu = 0.23$ and density $\rho = 3100$ \si{\kilo\gram\per\meter\cubed} are relatively well-known, and the fabrication pre-stress is known to be \SI{1.0}{\giga\pascal}. The thermal properties reported in literature vary (see~\cite{Chien2018} and references therein), we take the specific heat $C_p = 700$~\si{\joule\per\kilo\gram\per\kelvin} and thermal expansion coefficient $\Upsilon = 2.3\times 10^{-6}$~\si{\per\kelvin}. Literature values for the thermal conductivity vary between $k = 0.34$~\si{\watt\per\meter\per\kelvin}~\cite{Alam2012} and $k = 20$~\si{\watt\per\meter\per\kelvin}~\cite{Larsen2013}. We simulate using both values and plot the results as solid ($k = 20$~\si{\watt\per\meter\per\kelvin}) or dashed lines ($k = 2$~\si{\watt\per\meter\per\kelvin}) in Fig.~\ref{PhCthermal}\textbf{b,c,d}.

The simulation results show that for both thermal conductivities, the temperature (Fig.~\ref{PhCthermal}\textbf{b}) reaches a steady state after $\lesssim 0.3$\si{\second}, though the temperature is significantly higher for the low thermal conductivity case (\SI{470}{\kelvin}) than for the high thermal conductivity case (\SI{349}{\kelvin}). Due to this, the material expands by 20 (45)~\si{\nano\meter} for the high (low) thermal conductivity case ((Fig.~\ref{PhCthermal}\textbf{c}). On a lateral dimension of \SI{150}{\micro\meter} of the membrane this is negligible as it corresponds to approximately a 0.2 (0.4) \si{\nano\meter} increase in the lattice spacing $a$. Because of the width of the pad (middle of membrane) with respect to the tether, the stress in the pad is strongly reduced from the initial film stress (\SI{1}{\giga\pascal}), to $<20$~\si{\mega\pascal}. By the heating, this can be reduced to close to zero stress (compare yellow-dashed line in Fig.~\ref{PhCthermal}\textbf{d} to the blue line). Overall, these changes are not so significant and unlikely to change the dynamics of the membrane by themselves.

\subsubsection{Fast thermal behavior}\label{Fastthermal}
\begin{figure}
\includegraphics[width = 0.5\textwidth]{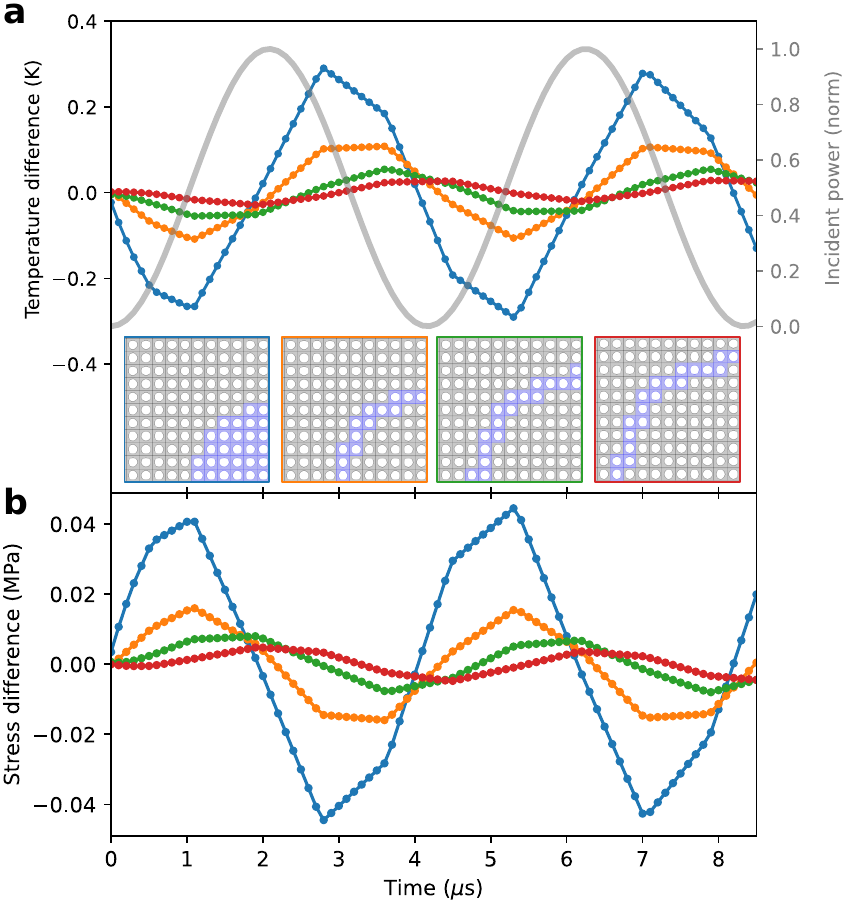}
\caption{\textbf{Fast thermal dynamics. a:} Incident laser power (grey) and average temperature difference with respect to steady state ($\simeq 348$~\si{\kelvin} for \SI{120}{\kilo\hertz} motion through standing wave optical trap field. Colors represent integration over domains depicted in the insets. \textbf{b:} Resulting stress change with respect to steady state.}
\label{PhCthermalfast}
\end{figure}
After simulating the steady state in temperature, material deformation and stress described in the previous subsection, we will now include the time-modulated optical intensity. We assume that the resonator moves initially in its fundamental mode at \SI{120}{\kilo\hertz}, so the optical field intensity is modulated by the $\sin$ term of Eq.~1. We make sure to keep the time-averaged power the same as in our steady-state simulation. We simulate a full period, and plot the incident heat flux (grey) and resulting temperature averaged over circular domains at various distances from the center (colors) in Fig.~\ref{PhCthermalfast}\textbf{a}. 

The incident heat flux modulates the temperature of the membrane with an amplitude of about ~\SI{0.3}{\kelvin} and a delay on the order of a microsecond. This delay is smallest in the domain where the heat flux is directly incident (blue), and increases as we move further away from the center (orange through red). The change in temperature also decreases as we move further away, which indicates that the effect is local to the laser beam spot. We can also extract the average von Mises stress, Fig.~\ref{PhCthermalfast}\textbf{b}, according to $\sigma_\mathrm{avg} = 1/\sqrt{2} \sqrt{(\sigma_{xx} - \sigma_{yy})^2 + 6\sigma_{xy}}$, with $\sigma_{xx,yy,xy}$ representing the stress components in the respective coordinates. This shows a clear inverse correlation with the temperature, and follows the trend of decreasing in amplitude and increasing in delay as we move further away from the incident laser beam spot. 

The resulting stress difference is relatively small compared to the total local steady-state stress ($\sim 24$~\si{\mega\pascal}). However, it can forms a parametric driving mechanism since stress is modulated at double the frequency of the original motion, guaranteed by the optical standing wave. This optothermal parametric driving is thus automatically frequency-matched to the dominant motion. The delay between heat flux and stress suggests that this mechanism of driving is limited in our case to $\lesssim 500$~\si{\kilo\hertz}, which corroborates with the fact that only observe combs based on the first few eigenmodes.

\subsubsection{Self-oscillation}\label{Selfoscillation}
\begin{figure}
\includegraphics[width = 0.5\textwidth]{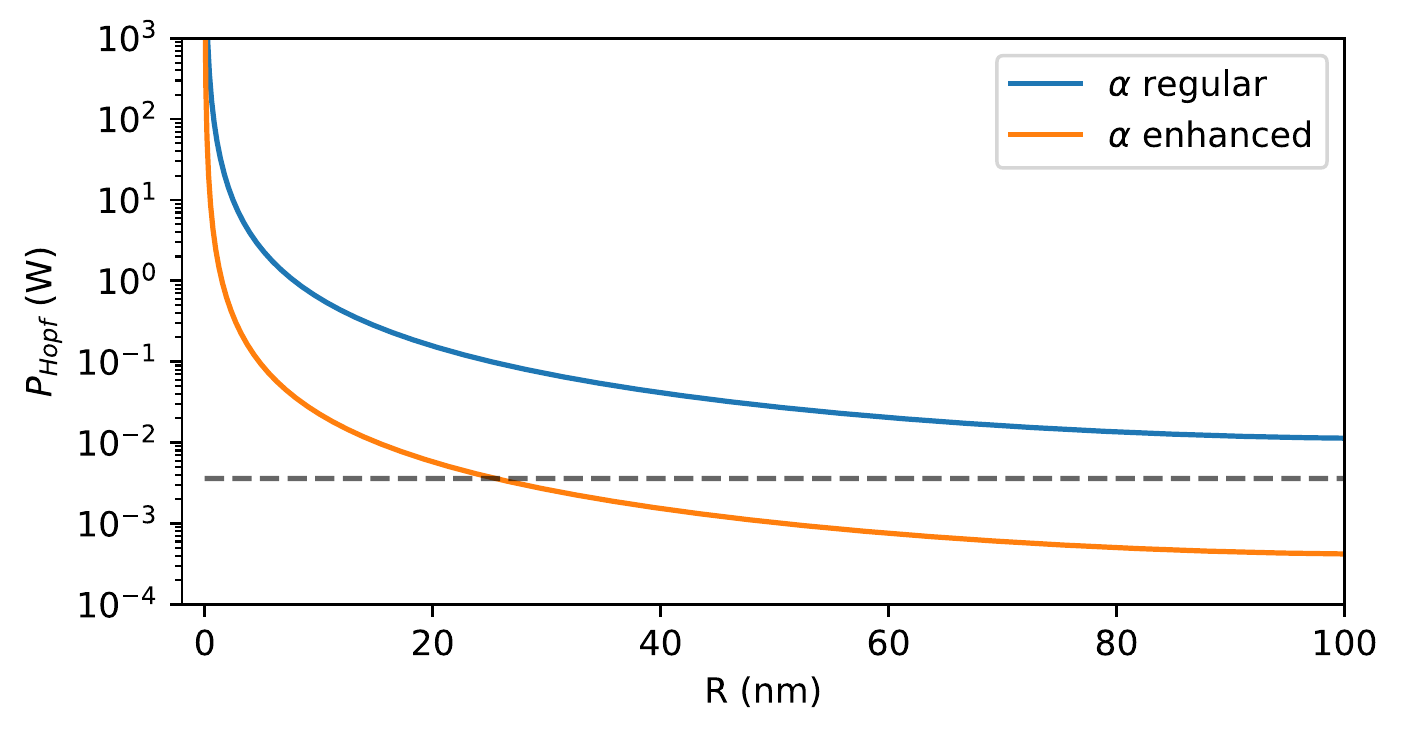}
\caption{Simulated optical power necessary to create a limit-cycle of amplitude $R$, using the absorption expected for a bare Si$_3$N$_4$ membrane, and with one enhanced by absorption through internal standing waves due to the photonic crystal. Black dashed lines indicate maximum laser power used in experiments, \SI{3.6}{\milli\watt}.}
\label{fighopfbifurcation}
\end{figure}
We have shown so far that the motion of the membrane through the optical standing wave results in a modulated stress via the absorption. However, this alone is not sufficient to drive the resonator to larger amplitudes since dissipation is also present (though small). To estimate whether this effect is strong enough to bring our resonator into limit-cycle oscillations under a $P\lesssim 3$~\si{\milli\watt} drive, we follow the model of \cite{Aubin2004}. There, the authors derive the minimum incident laser power to bring a resonator into the self-oscillation regime (Hopf bifurcation),
\begin{equation}
P_\mathrm{Hopf} = \frac{3(B^2+1)(B^2+4)}{2\pi^2 AQ\psi K},
\label{Hopfpower}
\end{equation}
where
\begin{equation}
\begin{aligned}
K &= \eta_1 \lambda^4 R^4 + \eta_2 \lambda^2 R^2 + \eta_3, \\
\eta_1 &= -4\pi^2(B^2+1)C, \\
\eta_2 &= -24\pi^2 (B^2+1)x_\mathrm{off}^2C/\lambda^2 - 24\pi^2(B^2+4)x_\mathrm{off}D/\lambda\\ 
&\quad+ 3(B^2+1)C, \\
\eta_3 &= -4D(B^2+4)(8\pi^2 x_\mathrm{off}^2/\lambda^2-3)x_\mathrm{off}/\lambda.
\end{aligned}
\end{equation}
Here, $A$ describes the heating due to the laser (units \si{\kelvin\per\watt}), $B$ is a dimensionless constant, $C$ represents the stiffness change due to temperature (units \si{\per\kelvin}), $D$ is the optothermal forcing term (units \si{\per\kelvin}), $Q$ is the mechanical Q-factor of our resonator, and $\psi$ is the fraction of optical field that forms a standing wave. These equations give the laser power necessary to create a limit-cycle of amplitude $R$.

The authors of \cite{Aubin2004} give helpful guide on how to estimate the parameters of Eq.~\eqref{Hopfpower} from the simulations of Secs.~\ref{Fastthermal} and \ref{Slowthermal}. To estimate the thermal parameters $A$ and $B$, we require the steady state temperature at the point of illumination (center) of Fig.~\ref{PhCthermal}, $T_\mathrm{dc}$ (difference with respect to the room temperature environment), and the amplitude of the temperature change due to the standing wave optical field from Fig.~\ref{PhCthermalfast}, $T_\mathrm{ac}$. Using
\begin{equation}
\begin{aligned}
A &= B T_\mathrm{dc} \\
B &= \frac{T_\mathrm{ac} \omega}{\sqrt{T_\mathrm{dc}^2 - T_\mathrm{ac}^2}}
\end{aligned}
\end{equation}
and the values of $T_\mathrm{dc}$ and $T_\mathrm{ac}$ reported in the previous subsections, we get $A = 222425$~\si{\kelvin\per\watt} (independent of thermal conductivity $k$), and $B = 1257$ ($k = 2$~\si{\watt\per\meter\per\kelvin}) or $B = 3971$ ($k = 20$~\si{\watt\per\meter\per\kelvin}).

The forcing parameters $C$ (parametric) and $D$ (direct) can be estimated from other simulations. By calculating the difference between eigenfrequencies of our structure at the minimum and maximum temperatures of Fig.~\ref{PhCthermalfast}, $\Delta \omega_0$ for the fundamental membrane mode, we can use $C = 2\Delta\omega_0 / (\omega_0 T_\mathrm{ac})$. We extract $\Delta \omega = 2\pi\times 200$~\si{\hertz} from a simulation where we impose only the stress change due to the thermal behavior, which gives $C\simeq 0.01$~\si{\per\kelvin} The parameter $D$ should be zero, because the tensile stress means that a temperature change does not lead to a direct out-of-plane displacement. This is true if the out-of-plane heat gradient is small enough, the tensile stress does not reach zero, and as long as our entire structure has the same thermal expansion coefficient (i.e. is from the same material). The latter is what distinguishes us from previous work on the optothermal excitation (bolometric backaction) driving a cantilever into self-oscillation~\cite{Metzger2008}, where the gold layer induces deflection under temperature change such that $D \neq 0$.

Finally, the we take mechanical quality factor $Q = 1\times 10^6$ and  $\psi = 1$. The latter we implicitly assumed already in Sec.~\ref{Fastthermal}, where the standing wave field cancels out fully (modulated power goes to zero). This is likely an overestimation due to losses and transmission through the Si backplane. The assumptions behind the derivation of \cite{Aubin2004} leading to Eq.~\eqref{Hopfpower} are somewhat different than for our system, since they use small displacement $x \ll \lambda$.

We evaluate Eq.~\eqref{Hopfpower} and plot the power necessary for a limit-cycle of radius $R$ in Fig.~\ref{fighopfbifurcation}. We distinguish between the absorption that we expect for an unpatterned Si$_3$N$_4$ film (1\% light absorption, blue), and if it would be enhanced a factor 3 by the standing wave (orange). For non-zero limit cycle amplitude $R$, the orange curve lies below the (maximum) power we send in, \SI{3.6}{\milli\watt} indicated by the black dashed line, which shows that our membrane can be driven to self-oscillation by the optothermal parametric effect. This requires a slightly higher absorption than would be expected based on the material parameters, which can be achieved due to the internal standing optical waves shown in the previous section. Fig.~\ref{fighopfbifurcation} shows that limit cycles with amplitudes larger than $R \simeq 30$~\si{\nano\meter} can exist. For $R\rightarrow 0$, no limit cycle appears which is due to $D = 0$ in our system. By evaluating the second derivative of $P$ with respect to $R$, $\mathrm{d}^2 P/\mathrm{d} R^2$, we find that the Hopf bifurcation is supercritical and thus our limit-cycle is stable~\cite{Aubin2004}.

The optothermal parametric drive forms an effective periodic modulation, which could also lead to a frequency comb, similar to the dielectrophoretic force. By including a periodic variation in the frequency $\omega_0$, we simulate this effect. The equation of motion is
\begin{equation}
\ddot{x} + \gamma \dot{x} + \omega_0^2 (1 + F_\mathrm{th} \sin{2\omega_0 t}) x = 0,
\label{ThermalEOM}
\end{equation}
such that the parametric term oscillates at twice the frequency of motion. With $F_\mathrm{th} = 1\cdot 10^{-5}$, we are close to the simulated frequency shift $\Delta \omega$. Numerically integrating Eq.~\eqref{ThermalEOM} yields an increase of velocity over time (even in the presence of damping $\gamma/2\pi = 0.2$~\si{\hertz}). However, the higher harmonics generated by the inclusion of this parametric drive are negligible, \SI{60}{\decibel} lower than the fundamental mode. Higher harmonics (above $2\omega_0$) do not appear visibly in the spectrum. While the optothermal parametric drive is sufficiently strong to drive the membranes to self-oscillation, it is not strong enough to result in a frequency comb without the dielectrophoretic force.

To conclude this section, we have simulated the optical and thermal behavior of our resonators and analyzed the possibility of self-oscillation. We find that the motion of the membrane through the standing wave optical field modulates the absorbed light intensity, which in turn modulates the tensile stress in the material. This is sufficiently strong to drive our resonator to self-oscillation, if the absorption is slightly higher than what is expected based on purely the material parameters. 

\subsection{Overtone comb with piezo driving}\label{Combwithpiezodriving}
\begin{figure}
\includegraphics[width = 0.5\textwidth]{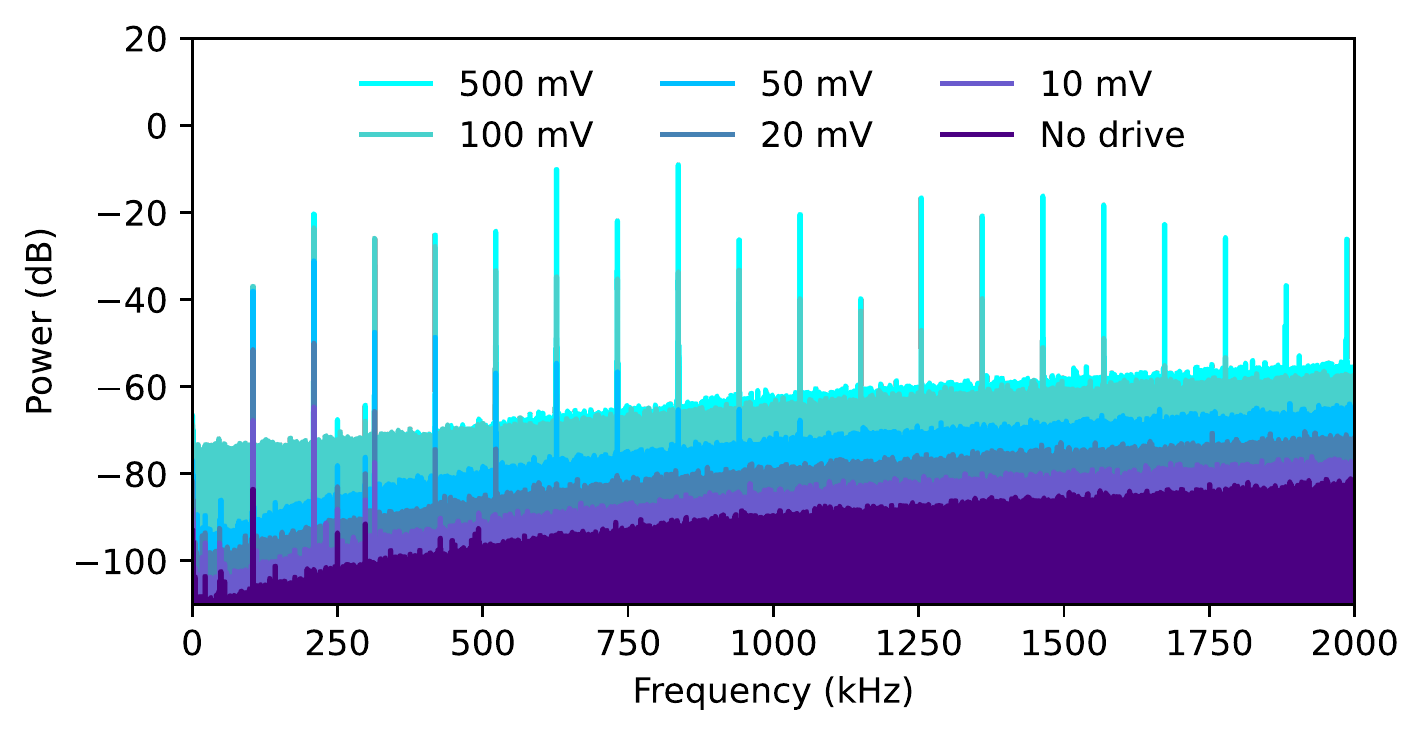}
\caption{\textbf{Comb with sine drive.} Observed velocity power spectra for different drive powers of a \SI{104.524}{\kilo\hertz} sine wave. The overtones appear when the resonator is driven with high powers on resonance, such that the displacement amplitude is large. The individual traces are offset by \SI{5}{\decibel} vertically.}
\label{Combwithsinedrive}
\end{figure}
In this section, we demonstrate that it is possible to generate the overtone frequency comb through inertial driving with a piezo shaker. We do this by using several membranes that do not show any sign of overtones when the laser is positioned anywhere on the membrane for at least five minutes. This way, we exclude thermal parametric driving. The piezo shaker is mounted on the back-side of the sample holder and driven via an Agilent 33220A arbitrary waveform generator and a a FLC A400 voltage amplifier with $20\times$ gain.

In Fig.~\ref{Combwithsinedrive}, we show the velocity power spectrum of a device that is driven on resonance by a sine wave of varying power. Without any drive, the fundamental mode (\SI{104.524}{\kilo\hertz}) can be observed, together with the second and third modes above \SI{250}{\kilo\hertz}. At higher powers, the power in the first mode grows and overtones appear. For the highest drive powers, we retrieve a comb similar to the one reported in the main text. This demonstrates that using an external drive tone, we can utilize the overtone mechanism to generate a frequency comb.

\begin{figure*}[t]
\includegraphics[width = \textwidth]{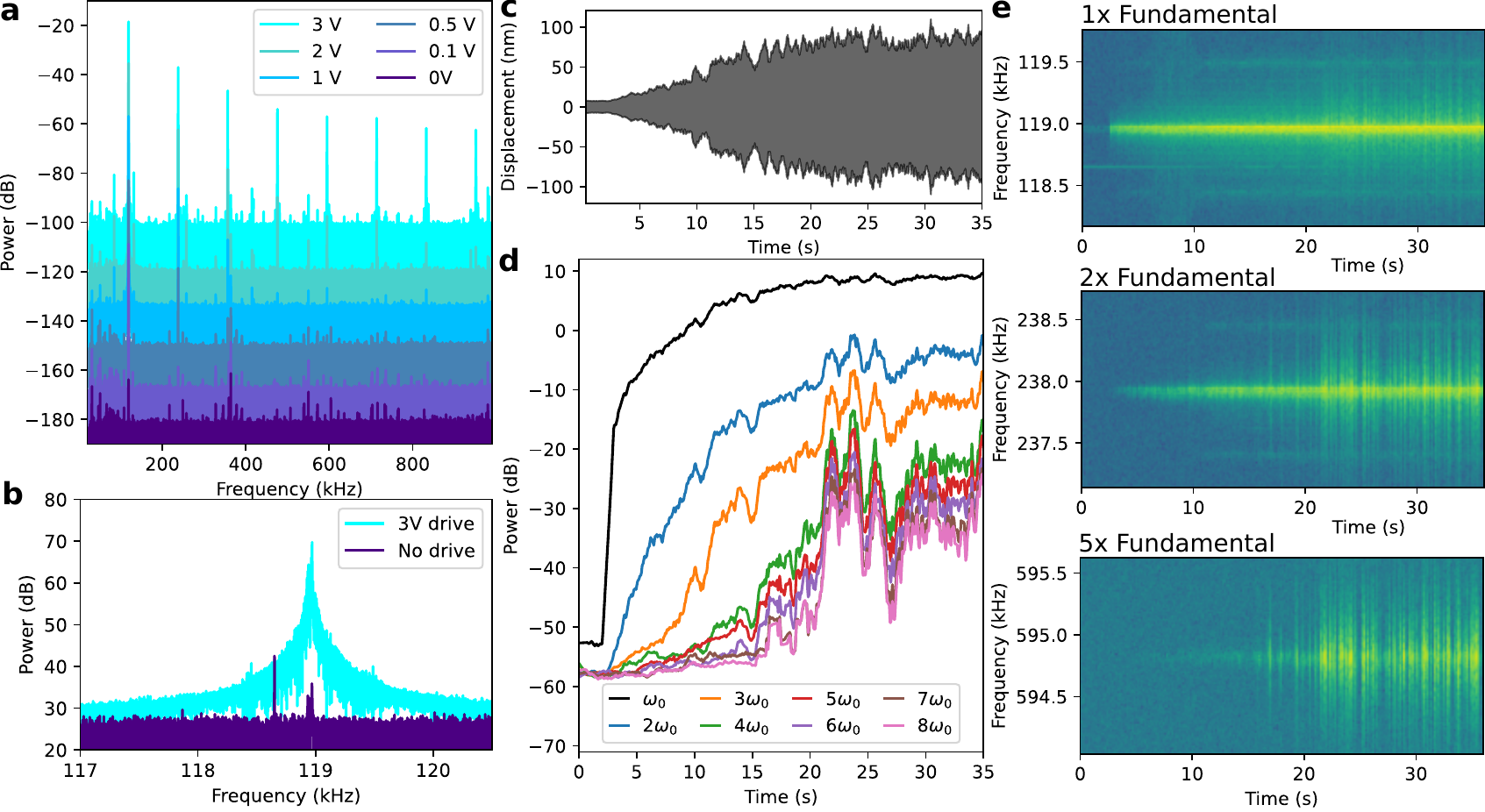}
\caption{\textbf{Overtone frequency comb through piezo driving. a:} Displacement power spectrum for various white noise driving powers. The overtone frequency comb only appears for the highest driving powers. \textbf{b:} Fundamental mechanical mode without and with driving. \textbf{c:} Displacement of the resonator as the drive power is slowly increased from $0$ to \SI{3}{\volt}. The displacement shows a less smooth signal than those obtained through thermal parametric driving in the main text. \textbf{d:} Extracted strengths of the individual comb teeth over time from the signal shown in \textbf{c}. Jumps around \SI{25}{\second} likely originate from the waveform generator internal switches. \textbf{e:} spectral maps of fundamental and first two overtones. The jumps observed in \textbf{d} cause the vertical lines.}
\label{figNoisedriving}
\end{figure*}

If the comb can be generated with the piezo by applying a white noise drive, it would be possible to avoid an external frequency reference that must be matched to the device. In Fig.~\ref{figNoisedriving}\textbf{a}, we apply a white noise drive of varying peak-to-peak amplitude. Again, the overtone comb is absent for lower drive powers, but clearly present once the drive is sufficiently strong. Fig.~\ref{figNoisedriving}\textbf{b} shows the fundamental mode for the lowest and highest drive cases.
We perform a similar analysis of the white-noise driven comb dynamics as for the optothermal parametrically driven comb in the main text. The displacement of the mechanical resonator over time as we increase the drive power is shown in Fig.~\ref{figNoisedriving}\textbf{c}, starting from \SI{0}{\volt} at $t\leq 2.5$~\si{\second} to \SI{3}{\volt} at $t\geq 25$~\si{\second}. Compared to those obtained via thermal parametric driving, the white noise drive results in a less smooth displacement signal. The amplitude is also noticeably smaller, which may be related to using a different displacement encoder, which is not impedance-matched to the oscilloscope used to record the time signal. By extracting the different overtones, Fig.~\ref{figNoisedriving}\textbf{d}, we see the fundamental mode increase in power first, the first and second overtone grow some seconds later and the higher overtones only appear once the amplitude has grown sufficiently. This matches the behavior of the comb shown in the main text. The bumps between $t=20$~\si{\second} and $t=28$~\si{\second} are due to the waveform generator internal switches. 

We have shown that we can generate the overtone frequency comb using different driving mechanisms. Using inertial driving via a piezoelectric shaker, we observe qualitatively the same comb behavior as when using the thermal parametric drive. However, the white noise drive results in a noisier displacement signal and comb teeth powers. To highlight this, we show the displacement power spectrum around the first three tones in Fig.~\ref{figNoisedriving}\textbf{e}, where the vertical lines indicate fast changes of comb power. This, combined with the added complexity of requiring a piezoelectric shaker, voltage source and amplifier, motivates the choice of using thermal parametric driving for the majority of the measurements in this work.

\subsection{Extension via comb interactions}\label{Seccombinteractions}
\begin{figure}
\includegraphics[width = 0.5\textwidth]{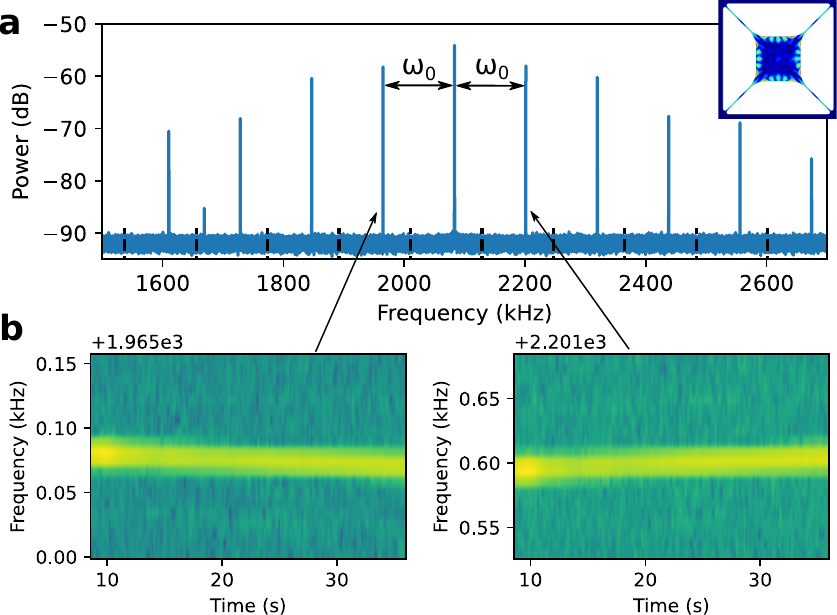}
\caption{\textbf{Comb interactions. a}: Spectrum around a higher-order mechanical mode, showing a symmetrized copy of frequency comb of the fundamental mode, $\omega_0$. The teeth are offset from the integer multiples of the original mode (dashed lines), but still equally spaced by $\omega_0$. Inset shows mode shape of the (5,5) mode expected to be the center peak at \SI{2083.3}{\kilo\hertz}. Readout position is identical to Fig.~2\textbf{c} of the main text, where the (5,5) mode has considerable amplitude. \textbf{b}: The teeth of the secondary comb follow the \SI{9}{\hertz} shift of the fundamental comb mode as in Fig.~3, and shift symmetrically away from the center peak.}
\label{Combinteractions}
\end{figure}
The bandwidth (span) of a frequency comb is an important property for many applications. While the overtone comb already performs on-par with the largest bandwidth mechanical combs (Table~\ref{Comparisoncombstable}), we illustrate how the bandwidth can be extended further. This can be done by letting the frequency comb interact with a higher-order mechanical eigenmode $\omega_\mathrm{h}$ of the membrane, to generate a comb with frequency spacing $\omega_0$ centered around $\omega_\mathrm{h}$. This comb around $\omega_\mathrm{h}$ has harmonics appearing on both sides of $\omega_\mathrm{h}$. In contrast, the comb of $\omega_0$ has harmonics only on one side. This means that the interaction with $\omega_\mathrm{h}$ could double the span.

We observe a copy of the fundamental mode overtone comb ($\omega_0$ spacing) symmetrically around the higher-order mode of the system. In Fig.~\ref{Combinteractions}\textbf{a}, we plot the spectrum around a higher order at $\omega_\mathrm{h} =$~\SI{2083.3}{\kilo\hertz} which we designate as the (5,5) mode shown in the inset. The simulated frequency of the (5,5) mode is at \SI{2046}{\kilo\hertz}, and this is the closest mode with significant mode amplitude at the readout position (same as in Fig.~2\textbf{c} of the main text). The symmetric pattern of $\omega_0$-spaced comb teeth is offset from the overtones of $\omega_0$ (black dashed lines).

To corroborate this interaction between the frequency comb as a whole and a higher-order mode of the membrane, we closely study the frequency of two of the comb teeth over time. Originating from the same measurement as Fig.~3 of the main text, the fundamental mode at $\omega_0$ has a \SI{9}{\hertz} upwards frequency shift over the duration of the measurement. In Fig.~\ref{Combinteractions}\textbf{b}, we plot the spectrum of two of the comb teeth (the first ones symmetrically around the central peak). These peaks show the same shift in frequency, but one shifts upwards while the other shifts downwards in frequency; they are clearly spaced around the central (5,5) mode peak by $\omega_0$. The teeth further away from the center follow the $9n$~\si{\hertz} scaling of the fundamental mode comb as described in the main text. This interaction thus unlocks the possibility of shifting the overtone comb upwards in frequency and thereby extend its span.  

\subsection{Effect of optics on comb and membrane}\label{Comblaser}
In this section, we demonstrate the overtone frequency comb on multiple devices to corroborate the measurements shown in the main text. Additionally, we also describe the visible changes in the membranes under an optical microscope when driven into the overtone comb regime, and we study the effect of the optics (power and focus) on the frequency comb.

\subsubsection{Comb prevalence and visible deformation}
\begin{figure}[!ht]
\includegraphics[width = 0.5\textwidth]{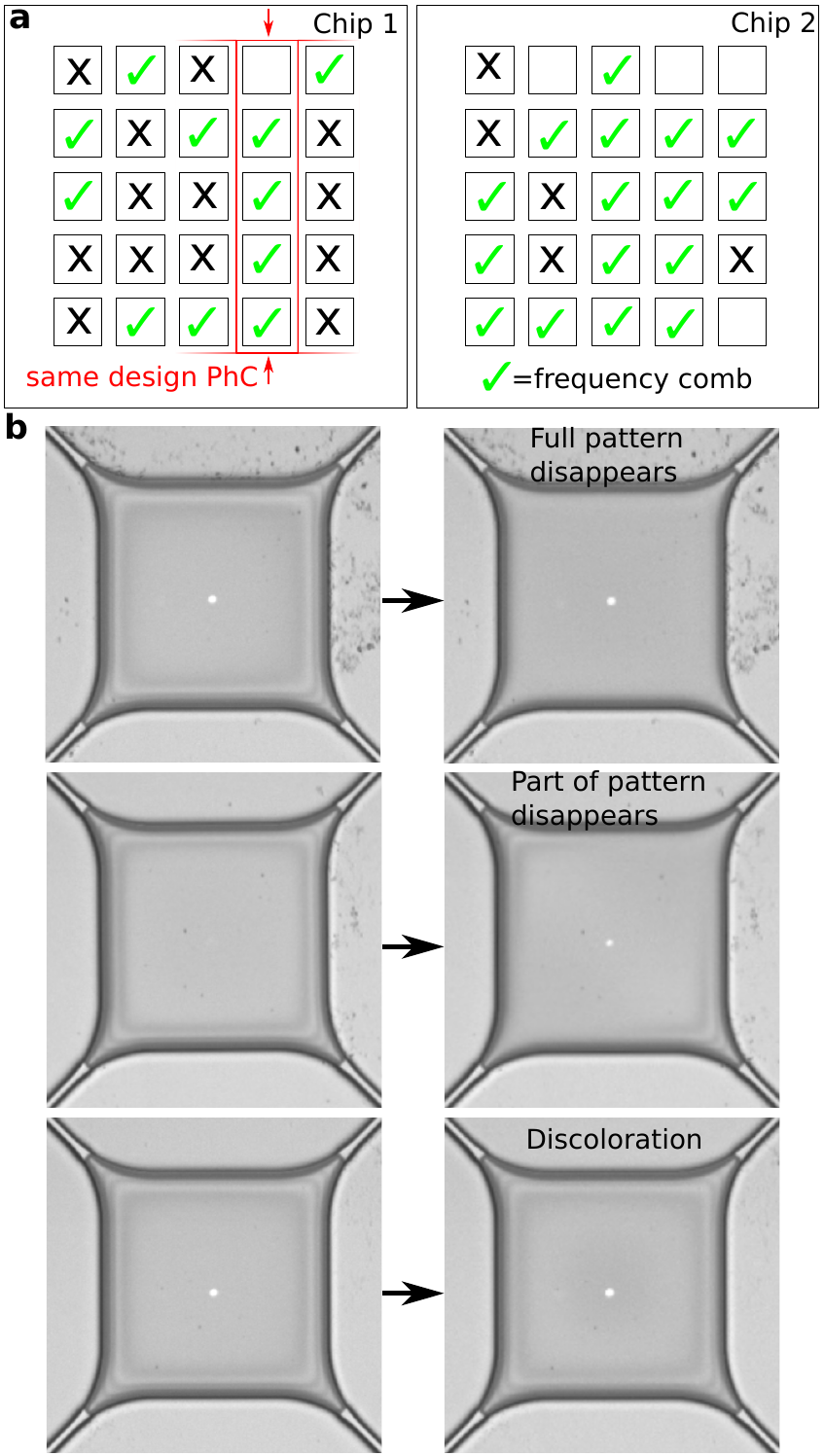}
\caption{\textbf{Visual changes in membrane structure. a:} Schematic sample maps of two nominally identical chips containing $25$ membrane devices each, using $5$ different photonic crystal designs. Green checkmarks indicate which devices show frequency comb behavior when optically addressed, blank squares show membranes collapsed during fabrication. \textbf{b}: Microscope images of three different devices before (left) and during (right) frequency comb behavior, showing visible changes. The color pattern along the outer edge of the membrane either disappears fully (top), partially (middle) or shows as a slight discoloration centered around the laser spot (bottom), indicating induced change in the photonic crystal lattice spacing and hole radius.}
\label{Structuralchange}
\end{figure}
We observe the overtone frequency comb behavior on multiple devices, shown in Fig.~\ref{Structuralchange}\textbf{a}. We fabricated two nominally identical chips containing $25$ membranes each. The membranes are identical with the exception of the photonic crystal parameters, as described in earlier work \cite{deJong2022b}. All of the designs feature the same internal modes that allow for strong absorption of $\lambda = 633$~\si{\nano\meter} light as in Sec.~\ref{Thermalparametricdriving}, albeit at slightly different wavelengths. Each chip contains the same set of $5$ photonic crystal designs, oriented as marked in Fig.~\ref{Structuralchange}\textbf{a}. 

Of the $45$ surviving devices on the two chips, $27$ demonstrated the frequency comb behavior based on optothermal parametric driving (Sec.~\ref{Thermalparametricdriving}), and the remainder could be driven into the overtone comb regime by sufficient piezoelectric shaking (Sec.~\ref{Combwithpiezodriving}). This confirms the generality of the overtone mechanism.  

We can corroborate the relation of the overtone comb and (powerful) heating through absorption by studying the membranes using the microscope of the laser Doppler vibrometer. The presence of the optothermal parametrically driven frequency comb behavior is correlated with visible changes in the membrane structure. We show this by taking a microscope image immediately after the laser spot is moved to the center of the membrane (Fig.~\ref{Structuralchange}\textbf{b}, left column), and $>30$~\si{\second} later when the system has reached steady state (Fig.~\ref{Structuralchange}\textbf{b}, right column). There is an interference pattern at the edges of the membrane, which is likely from the change in photonic crystal pattern due to a stress-gradient from the edge of the Si$_3$N$_4$. The laser beam heats the membrane and via the thermal expansion of the Si$_3$N$_4$, the stress changes such that this edge deforms and the interference pattern changes. 

\begin{figure}
\includegraphics[width = 0.5\textwidth]{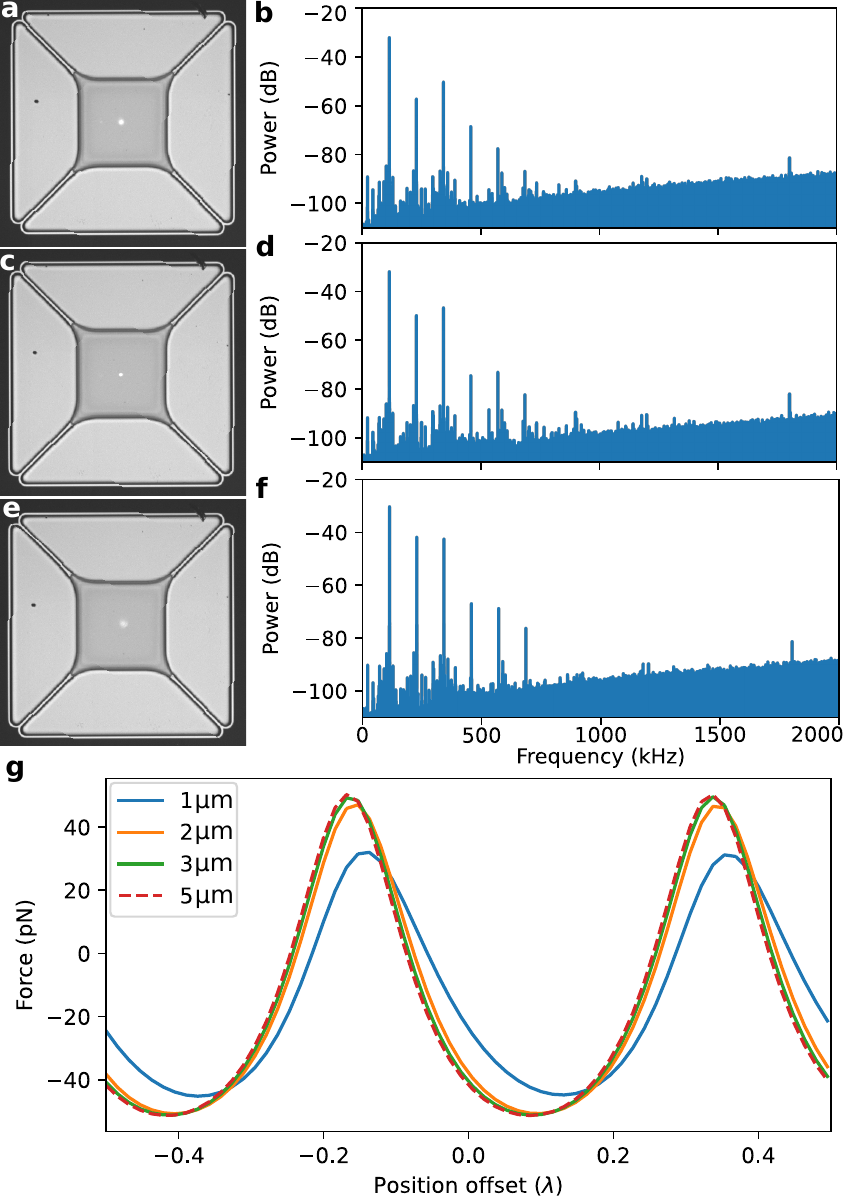}
\caption{\textbf{Comb dependence on laser focus. a,c,e:} Microscope images of the same device with the laser focus before (\textbf{a}), at (\textbf{c}) and after (\textbf{e}) the plane of the membrane structure. \textbf{b,d,f:} Frequency comb spectra measured with the respective laser foci, showing no major change. \textbf{g:} Simulated trapping force as a function of beam waist for constant total beam power.}
\label{Focuschange}
\end{figure}

We show three devices to illustrate different changes observable in this interference pattern in Fig.~\ref{Structuralchange}\textbf{b}. In some devices (top row), the pattern disappears completely, which suggests these have the strongest absorption: The deformation is larger than in other devices and the steady-state temperature is thus high. Simultaneously, these devices show strong frequency combs. In other devices (Fig.~\ref{Structuralchange}\textbf{b}, middle row), the interference pattern disappears only partially. This is typically along one diagonal, which suggests that the stress along the other diagonal is not fully removed; we can switch between the diagonals by positioning our laser spot. The steady-state temperature is likely less than in the case where the pattern disappears completely, and also the frequency comb appears less pronounced. Finally, a few devices show a small discoloration around the laser spot, which suggests that the interference pattern is modified only slightly. These devices display only weak frequency combs. These behaviors corroborate that the thermal parametric driving through absorption of the light is the origin of the frequency combs that we observe.

\subsubsection{Comb dependence on optical focus}
We study the effect of laser focus on the frequency comb behavior. By deliberately defocusing the laser beam, and using a membrane that has a weak frequency comb behavior, we expect the focus to change the number of overtones appearing in the spectrum. 

In Fig.~\ref{Focuschange}\textbf{a,c,e}, we show microscope images of the laser focus before, at and after the plane of the membrane. In \textbf{b,d,f}, we plot the velocity power spectrum measured from the devices with their respective foci, which are nearly identical. This shows that the formation of the frequency comb does not depend significantly on the laser focusing. Simulations of the optical trapping force as a function of beam waist (Fig.~\ref{Focuschange}\textbf{g}) confirm this behavior: The trapping force does not change for beam waists $> 3$~\si{\micro\meter}. Simultaneously, the thermal parametric force is limited to the area of the beam spot, so as long as the total beam power is constant, the changes in power density and beam area will cancel and the thermal parametric driving will be constant. This shows that the generation of an overtone frequency comb does not require a tightly focused laser beam.  

\subsubsection{Comb dependence on optical power}
\begin{figure}
\includegraphics[width = 0.50\textwidth]{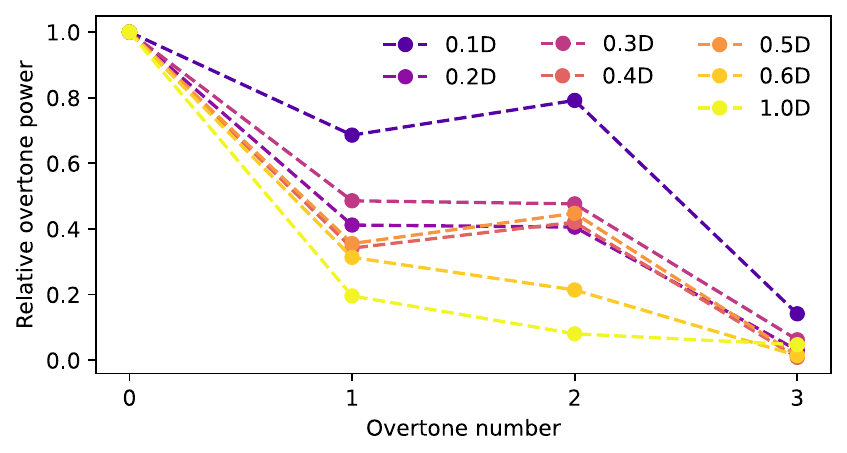}
\caption{\textbf{Comb dependence on optical power.} Relative powers in the individual overtones, normalized the overtone powers without filter and to the power of the fundamental mode for each filter setting. With less optical power, the overtones decrease in power with respect to the fundamental mode, as simulated in Fig.~\ref{fignumericalsim}. The labeled optical densities correspond to a transmission of $0.1\mathrm{D}: 79\%$, $0.2\mathrm{D}: 63\%$, $0.3\mathrm{D}: 50\%$, $0.4\mathrm{D}: 32\%$, $0.5\mathrm{D}: 32\%$, $0.6\mathrm{D}: 25\%$, $1.0\mathrm{D}: 10\%$.}
\label{Objectivechange}
\end{figure}
We also study the effect of laser power on the frequency comb. We can add an absorptive neutral density filter to reduce the power of the optical beam. This decreased $F_\mathrm{o}$ and should decrease the power of the overtones with respect to the fundamental mode, as simulated in Fig.~\ref{fignumericalsim}. In Fig.~\ref{Objectivechange}, we show the relative overtone powers measured on a single device, for various filter strengths. The  power in every overtone is normalized to the overtone power for the measurement without optical filter. Subsequently, the powers of the individual filter traces are then normalized for the power in their respective fundamental mode (overtone number $0$). $F_\mathrm{o}$ affects the power relative to the fundamental mode, not the absolute power.

From Fig.~\ref{Objectivechange}, we see a trend where the higher optical density filters lead to lower relative overtone powers. This matches the simulations shown in Fig.~\ref{fignumericalsim}. For higher density filters and higher overtone numbers ($n\geq 4$), the signal quality is too poor extract a meaningful relative overtone power. Nonetheless, it is a clear validation of the model of Eq.~\eqref{SIEOM} as the correct description of the source of the frequency combs.

\subsection{Comb stability}\label{Combstability}
\begin{figure*}
\centering
\includegraphics[width = 0.65\textwidth]{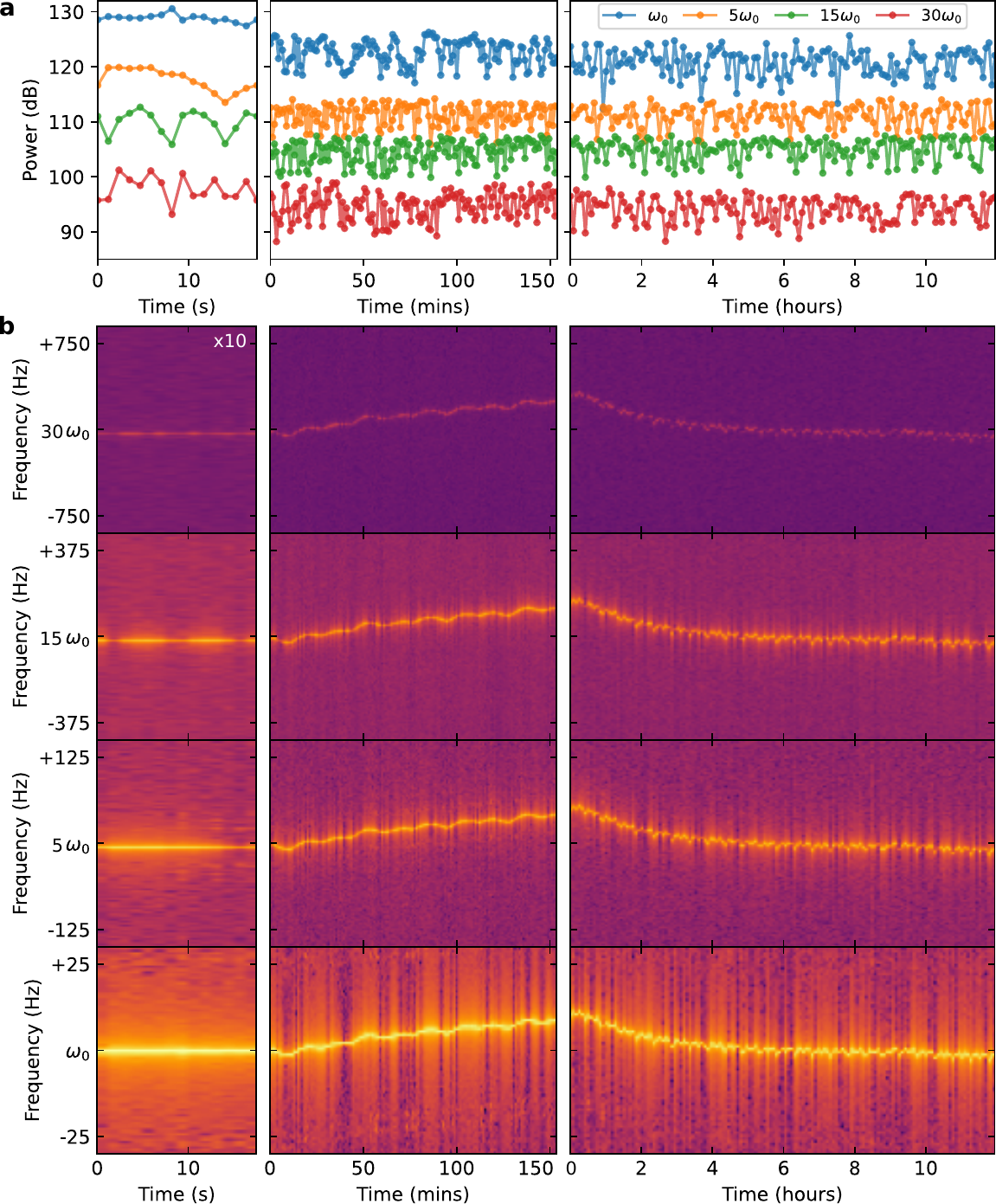}
\caption{\textbf{Comb stability. a}: Extracted power of different overtones ($\omega_0$, $5\omega_0$, $15\omega_0$ and $30\omega_0$, measured on the same device for several timescales. There is no correlation in the amplitudes on timescales longer than a minute, but there appears to be a beating pattern on short timescales. \textbf{b:} Frequency of different overtones, measured for several timescales. The frequency is stable on short timescales, but drifts due to temperature on timescales longer than a minute. The drift is the same (scaled by overtone number) for all overtones.}
\label{Figcombstability}
\end{figure*}
An important property of a frequency comb is the stability of the comb teeth over time, both in amplitude and in frequency. To gain insight in how our overtone frequency comb performs, we park our laser on a single device and measure the spectrum on various timescales. From each spectrum, extract the frequency and amplitude of some of the overtones, and plot them in Fig.~\ref{Figcombstability}. 

In Fig.~\ref{Figcombstability}\textbf{a}, we plot the power of the overtones at $\omega_0$, $5\omega_0$, $15 \omega0$ and $30\omega_0$, offset vertically for clarity. On the timescales $>1$~\si{\minute}, the amplitude of the individual overtones fluctuates around a constant mean with a 95\% confidence interval between \SI{2.0}{\decibel} and \SI{2.5}{\decibel}. On shorter timescales, a correlation between the amplitudes is visible which can be reproduced by simulations if decay is included. 

In Fig.~\ref{Figcombstability}\textbf{b}, we show the frequency of the same set of overtones, where the width of the displayed part of the spectrum is proportional to the overtone number. The color indicates power and is the same for the bottom three rows, but multiplied by a factor 10 for the top row for clarity. All comb teeth display the same frequency shift over time, scaled by overtone number. This correspondence is \emph{exact}, to within the resolution of our measurement. On short timescales ($< 16$~\si{\second}), the frequency of the fundamental mode is stable to within the resolution of our measurement (\SI{0.78}{\hertz}). However, we can use the exact frequency correspondence and observe a slight shift of the overtone at $30\omega_0$, \SI{4.68}{\hertz} over the measurement time, or \SI{7.9e-8}{\per\second} relative frequency stability. This is likely limited by the environment temperature causing a frequency shift of the mechanical mode. That shift is clearly visible as the slow drift on longer timescales, as the right-most column was measured overnight and the frequency seems to stabilize. There are also oscillations on a~\SI{10}{\minute} timescale, which we attribute to the room airconditioning. If we select the most stable 6-hour period of our overnight measurement, the fractional stability reaches \SI{7.5e-10}{\per\second}.

\subsection{Noise in overtone frequency combs}\label{Phasenoisederivation}
We study fluctuations in the frequency comb as a response to (thermal) noise. By starting from Eq.~1 and including a Langevin force noise term $\xi(t)$, we obtain
\begin{equation}
\ddot{x} + \gamma \dot{x} + \omega_0^2 x = F_\mathrm{o} \sin\left(\frac{4\pi}{\lambda}(x-x_\mathrm{off})\right) + F_\mathrm{d} e^{i\omega_d t} + \xi(t),
\label{EOMnoise}
\end{equation}
Where the noise has zero mean $\langle \xi(t) \rangle = 0$ and is correlated $\langle \xi(t) \xi(t') \rangle = 2\gamma T \delta(t-t')$ such that we have assumed Markovian noise at temperature $T$, valid since resonator $Q = \frac{\omega_0}{\gamma} \gg 1$~\cite{Kac1981}. We expand the sine-squared nonlinear term in the same way as in Sec.~\ref{Overtones}, where we keep all the terms since the displacement is not small. 
We can rewrite this to
\begin{equation}
\ddot{x} + \gamma \dot{x} + \omega_0^2 x = \sum_n^\infty \alpha_n x^n + F_\mathrm{d} e^{i\omega_d t} + \xi(t),
\label{EOMnoisereduced}
\end{equation}
such that if $x_\mathrm{off}= 0$, all odd coefficients ($\alpha_1,~\alpha_3,~\alpha_5,$...) are zero. Let us first consider the oscillation of only the first harmonic term, $x =\frac{1}{2} \left( A e^{i\omega_0 t} + A^* e^{-i\omega_0 t}\right)$. Then expand the series of $\alpha_n x^n$ and truncate all the tones that are not resonant. The first few terms read
\begin{equation}
\begin{aligned}
\alpha_1 x &= \frac{\alpha_1}{2}\left( Ae^{i\omega_d t} + A^*e^{-i\omega_d t} \right)\Rightarrow \alpha_1 x \\ 
\alpha_2 x &= \frac{\alpha_2}{4}\left( A^2 e^{2i\omega_d t} + 2|A|^2 + A^{*2} e^{-2i\omega_d t} \right) \Rightarrow 0 \\
\alpha_3 x &= \frac{\alpha_3}{8}\left(\splitfrac{A^3 e^{3i\omega_d t} + 3|A|^2Ae^{i\omega_d t}}{+3|A|^2A^* e^{-i\omega_d t} + A^{*3}e^{-3i\omega_d t}} \right) \Rightarrow \frac{3\alpha_3}{4}|A|^2 x \\
\alpha_4 x &= \frac{\alpha_4}{16} \left( \splitfrac{A^4 e^{4i\omega_d t} + 4|A|^2 A^2 e^{2i\omega_d t} + 6|A|^4}{+ 4|A|^2 A^{*2} e^{-2i\omega_d t} + A^{*4}e^{-4i\omega_d t}} \right) \Rightarrow 0.
\end{aligned}
\end{equation}
We can thus rewrite Eq.~\eqref{EOMnoisereduced} as
\begin{multline}
\ddot{x} + \gamma \dot{x} + \omega_0^2 x = \left[ \alpha_1 x + 0 + \frac{3\alpha_3}{4}|A|^2 x + 0 + \frac{5\alpha_5}{8}|A|^4 x + ...\right]\\ + F_\mathrm{d} e^{i\omega_d t} + \xi(t).
\end{multline}
Now we can introduce a new set of coefficients $\beta_n$ such that $\beta_1 = 1,~\beta_2 = 0,~\beta_3 = \frac{3}{4},~\beta_4 = 0,~\beta_5 = \frac{5}{8}$. Rearranging the terms now gives us
\begin{equation}
\begin{aligned}
\ddot{x} + \gamma \dot{x} + \left[1 - \sum_n \alpha_n \beta_n |A|^{n-1}\right]\omega_0^2 x &= F_\mathrm{d} e^{i\omega_d t} + \xi(t) \\
\ddot{x} + \gamma \dot{x} + \Omega_0^2 x &= F_\mathrm{d} e^{i\omega_d t} + \xi(t).
\end{aligned}
\end{equation}
This way, we work with the frequency of the first harmonic, $\Omega_0$, taking into account the shift due to the non-linearity. In the case where $x_\mathrm{off}= 0$, all odd $\alpha_n=0$, and all even $\beta_n = 0$ in general, which implies that the frequency of the first harmonic is not shifted by the non-linearity. Intuitively, that makes sense since the electric field is symmetric around the resonator rest position for $x_\mathrm{off} = 0$. 

We obtain the amplitude of the first harmonic by switching off the noise $\xi(t)$, and get
\begin{equation}
|A|^2 = \frac{F_d^2}{\left(\left[1 - \sum_n \alpha_n \beta_n |A|^{n-1}\right]\omega_0^2 -\omega_d^2\right)^2 + \gamma^2\omega_d^2}.
\end{equation}

Similarly, we can find the amplitude response to fluctuations $\xi(t)$ by turning off the coherent drive,
\begin{equation}
\ddot{x} + \gamma \dot{x} + \Omega_0^2 x = \xi(t).
\end{equation}
In Fourier space, we then get
\begin{equation}
X(\omega) = \frac{\xi(\omega)}{\Omega_0^2 - \omega^2 + i\gamma\omega}
\end{equation}
with autocorrelation
\begin{equation}
\left\langle X(\omega) X(\omega') \right\rangle = \frac{2\gamma k_\mathrm{B}T \delta(\omega + \omega')}{(\Omega_0^2 - \omega^2 + i\gamma\omega)(\Omega_0^2 - \omega'^2 + i\gamma\omega')}.
\end{equation}
Then we can find the mean square amplitude response to fluctuations as 
\begin{equation}
\begin{aligned}
\overline{\delta x^2} &= \left\langle X(t)^2 \right\rangle = \int \frac{\mathrm{d}\omega}{2\pi} \frac{\mathrm{d}\omega'}{2\pi} \left\langle X(\omega) X(\omega') \right\rangle e^{i(\omega + \omega')t} \\
&= \frac{2k_\mathrm{B}T}{\Omega_0^2}.
\end{aligned}
\end{equation}
Thus the mean square amplitude of the first harmonic due to thermal fluctuations depends on the optical non-linearity via $\Omega_0^2$. 

Now to obtain the net motion we consider both driving and noise terms (use Eq.~\eqref{EOMnoisereduced}) and make the ansatz that our solution consists of a set of harmonics with a perturbation $\delta x$
\begin{equation}
x = \sum_{n=1}^\infty A_n e^{in\omega_0 t} + \sum_{n=1}^\infty A_n^* e^{-in\omega_0t} + \delta x = B + \delta x.
\end{equation}
The nonlinear term $\sum_n \alpha_n x^n$ from Eq.~\eqref{EOMnoisereduced} can then be expanded and simplified by neglecting higher-order noise terms and keeping only the linear one. If $\delta x \ll B$,  $(B + \delta x)^n \simeq B^n + n B^{n-1} \delta x$. Now if we let the sum in the nonlinear term go to a reasonable finite number $N_\mathrm{max}$, we split the sum and shift the index of one of the two,
\begin{equation}
\begin{aligned}
\sum_{n=1}^\infty \alpha_n(B^n + nB^{n-1} \delta x) = \sum_{n=1}^{N_\mathrm{max}}\alpha_n B^n + \sum_{n=0}^{N_\mathrm{max}-1}\alpha_{n+1}(n+1)B^n\delta x& \\ 
= \alpha_1 \delta x + \sum_{n=1}^{N_\mathrm{max}-1} B^n(\alpha_n + \alpha_{n+1}(n+1)\delta x) + \alpha_{N_\mathrm{max}}B^{N_\mathrm{max}}&.
\end{aligned}
\end{equation}
If we then have $\alpha_n \simeq \alpha_{n+1}$ for all reasonable $n$, and small noise such that $\delta (n+1) \ll 1$, we can then absorb the term at $N_\mathrm{max}$ back into the sum,
\begin{equation}
\sum_{n=1}^\infty \alpha_n(B^n + nB^{n-1} \delta x) \simeq \alpha_1 \delta x + \sum_{n=1}^{N_\mathrm{max}} \alpha_n B^n.
\end{equation}

Thus we obtain the same solution in terms of the harmonics as we would do without considering fluctuations $\delta x$, with only the additional term $\alpha_1 \delta x$. The equation of motion thus becomes

\begin{widetext}
\begin{equation}
\sum_{n=1}^\infty A_n\left(\omega_0^2 - n^2 \omega_d + in\gamma\omega_d\right) e^{in\omega_dt} + \ddot{\delta x} + \gamma \dot{\delta x} + \omega_0^2 \delta x = \alpha_1 \delta x + \sum_{n=1}^{N_\mathrm{max}} \alpha_n B^n + F_d e^{i\omega_d t} + \xi(t)
\end{equation}
Based on the previous analysis, we can linearize this to
\begin{equation}
\sum_{n=1}^\infty A_n\left(\omega_0^2 - n^2 \omega_d + in\gamma\omega_d\right) e^{in\omega_dt} + \ddot{\delta x} + \gamma \dot{\delta x} + \Omega_0^2 \delta x + F_d e^{i\omega_d t} + \xi(t)
\end{equation}
\end{widetext}
where $\Omega_m^2 = \left[1 - \sum_n \alpha_n \beta_n |A|^{n-1}\right]\omega_0^2$. 
While the amplitudes $A_n$ are obtained in Sec.~\ref{Overtones}, this analysis shows that the root-mean square fluctuations stay constant at $\frac{2k_\mathrm{B}T}{\Omega_m^2}$ for all values of $n$. This indicates that the comb generation does not cause additional fluctuations or noise.

Thus we have confirmed our observations of Fig.~3\textbf{f} in the main text: The Lorentzian linewidth of our resonator fundamental mode is the same in the thermal regime as it is when driven into the overtone comb, since the optical non-linearity does not add noise to the system.

\subsection{Phase-coherence}\label{Secphasecoherence}
To show the phase-coherence of all tones within the comb, we take a closer look at the time domain signal. When the overtones at $2,3, 4, ... \times\omega_0$ have considerable amplitude, we can extract the phases of each component separately from the shape of a single period in the time domain. To do so, fit the sum of the first $n$ cosine terms with amplitude coefficients $a_1, ..., a_{n}$ and phase offsets $\phi_1, ..., \phi_{n}$ using
\begin{equation}
\begin{aligned}
A(t) &= a_1\cos(2\pi\omega_0 t + \phi_1) + a_2\cos(4\pi\omega_0 t + \phi_2) + ... \\
 &+ a_{16}\cos(32\pi\omega_0 t + \phi_{n}).
\label{eqphasecoherence}
\end{aligned}
\end{equation}
When all phase offsets are the same, $\phi_1 = \phi_2 = ... = \phi_{n}$, the overtone comb is phase-coherent. 

\begin{figure}
\includegraphics[width = 0.5\textwidth]{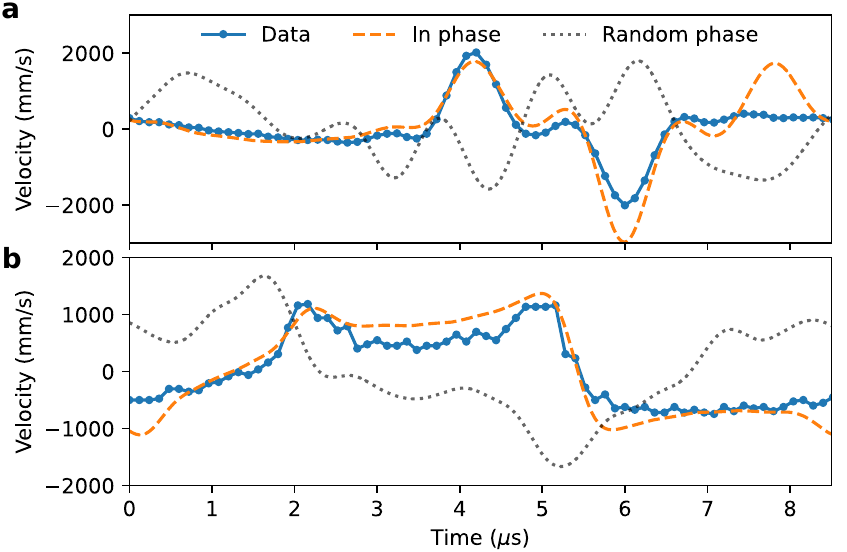}
\caption{\textbf{Phase coherence of overtone comb. a,b}: Single period (blue) of the measured time signals of Fig.~4\textbf{c}, together with fits containing the sum of the first 16 cosines (overtones, at $\omega_0$, ..., $16\omega_0$). In orange, the cosines all have the same phase offset, while in black the phase offset is random. This shows the phase-coherence of our overtone comb.}
\label{Figphasecoherence}
\end{figure}

To fit Eq.\eqref{eqphasecoherence}, we extract the amplitudes of the first $n=16$ overtones from the measurement of the whole time signal ($\sim 36$~\si{\second}). This we use as an initial guess for a fit that optimizes them to the final amplitudes that best describe a single period ($\sim 8$~\si{\micro\second}). We calculate two curves, one where all phase offsets $\phi_1, ..., \phi_{16}$ are the same (orange in Fig.~\ref{Figphasecoherence}\textbf{a,b}) and one where all the phase offsets are random (black). From this, it is clear that all overtones  have the same phase offet and thus the overtone comb is phase-coherent.

\end{document}